\documentclass[aps,superscriptaddress,floatfix,showpacs,letterpaper,nofootinbib]{revtex4}

\usepackage{graphicx}
\usepackage{enumerate}

\def\beq{\begin{eqnarray}}
\def\eeq{\end{eqnarray}}
\def\nnb{\nonumber}

\newcommand{\be}{\begin{equation}}
\newcommand{\ee}{\end{equation}}
\newcommand{\bea}{\begin{eqnarray}}
\newcommand{\eea}{\end{eqnarray}}
\newcommand{\ba}{\begin{array}}
\newcommand{\ea}{\end{array}}

\def\nnb{\nonumber}
\def\xslash\sharp1{{\rlap{$\sharp1$}/}}

\def\be{\begin{equation}}
\def\ee{\end{equation}}
\def\bea{\begin{eqnarray}}
\def\eea{\end{eqnarray}}

\def\nnb{\nonumber}

\begin{document}

%\begin{frontmatter}

\title{ Investigation of a kind of neutrino mass matrix}

\author{Chao-Shang Huang$^{1 \dag}$,
Wen-Jun Li$^{2 \ddag}$\vspace{5mm}\\
\normalsize \emph{
1.CAS Key Laboratory of Theoretical Physics, Institute of Theoretical Physics,}\\
\normalsize \emph{ Chinese Academy of Sciences, Beijing 100190, China}\\
\normalsize \emph{2.Physics school, Henan Normal
University, Xinxiang 453007, China}\\
\normalsize \emph{E-mail: $^{\dag}$csh@itp.ac.cn, $^{\ddag}$liwj24@163.com}}

%\begin{abstract}
%it will come later
%\end{abstract}
%\section{Symbol psyr}
%\fonttable{psyr}
%\maketitle

\begin{abstract}

\section*{Abstract}

We carry out diagonalization of a kind of Majorana neutrino mass matrix $M=M_\nu M_\nu^\dagger$ of which Real part and Imaginary part are commutative. For the kind of matrix M, it is shown in a model-independent way that $\delta = \pm\pi/2$ which implies the maximal strength of CP violation in neutrino oscillations and atmospherical mixing angle would be in the ranges, $\pi/4 < \theta_{23} < 3\pi/4$, or $-3\pi/4<\theta_{23} < - \pi/4$. It is shown that the kind of Hermitian Majorana neutrino mass matrix M has only five real parameters and furthermore, only one free real parameter (A or D) if using the measured values of three mixing angles and mass squared differences as inputs.

\end{abstract}

\maketitle
%\end{frontmatter}
%%%%%%%%%%%%%%%%%%%%%%% Main body %%%%%%%%%%%%%%%%%%

\section{Introduction}
 Massive data from solar, atmospheric, reactor and accelerator neutrino experiments facilitate research of neutrino mass and mixing in past thirty years. Knowledge of these magic particles has initially been grasped. However, the structure of
 neutrino mass matrix and their flavor mixing mechanism hidden behind are still not clear.

   To understand the flavor structure of neutrinos is one of the
outstanding problems in search for new physics. A series of attempts have been made to build the structure of
 neutrino mass matrix and their flavor mixing mechanism hidden behind.
Some have considered inserting flavor symmetries to the standard model(SM), such as Tribimaximal mixing(TBM)\cite{tbm},TM1 and TM2 mixings\cite{He}, texture zeros\cite{Xi}, vanishing cofactors\cite{van}, equalities\cite{eq}, hybrid textures\cite{hy} and $\mu-\tau$ symmetry\cite{king}. Recently,  the case of neutrino mass matrix with the predicted atmospheric neutrino mixing angle $\theta_{23}$ near $45^o$  in
type-I+II seesaw mechanism using non-Abelian flavor symmetry $A_4$ has been discussed\cite{RR}. The implementation of $U(1)_{L_{e}-L_{\tau}}$ gauge symmetry have been used to study the
neutrino phenomenology within the framework of type-(I+II) seesaw\cite{mkb}. Similar study with $S_3\bigotimes Z_2$ symmetry also has been finished\cite{JD}. In Ref.\cite{ai}, a specific texture characterized by one equality between two
independent elements in the neutrino mass matrix has been explored and it is noted that all the 15 possible one-Equality textures can accommodate the experimental data in the case of normal ordering, whereas only fourteen one-Equality patterns are viable in the inverted hierarchy type. Moreover, the authors in Ref.\cite{yu} have investigated the new texture of Majorana neutrino mass matrix under cosmic microwave background radiation and neutrinoless double beta decay measurements. Additionally, based on the cosmic microwave background temperature fluctuation and polarization
measurements, Supernovae Ia luminosity distances, Baryon Acoustic Oscillation observations and
determinations of the growth rate parameter, the most constraining bound have derived as $\sum m_\nu <0.09 eV$ at $95\%$ CL\cite{edv}. A model-independent way have been performed to constrain effective Majarana neutrino masses $|<M_{\alpha\beta}>|(\alpha,\beta=e, \mu,\tau)$ with current experiment data, and got a few information of the absolute neutrino mass scale and one of the effective Majorana CP phases\cite{Xi1}. In Ref.\cite{Xic}, it is shown that two Majorana CP violating phases could
be constrained from combination of neutrinoless double beta decay experimental constraint and cosmological constraint on effective Majorana neutrino mass. In addition, from the view of energy scales dependence,  the neutrino mixing parameters under simple, ultraviolet-complete models have been discussed and several robust experimental signatures, some observable effects on neutrino experiments have been shown\cite{babu}. About the existing neutrino experiment factories, the authors in Ref.\cite{ab} have made an analysis of their systematic uncertainty and physics reach, and conceived a new generation of short-baseline cross section experiments.

Assuming that neutrino is Majorana fermion, one can express flavor eigen-states in terms of mass eigen-states: \\
\bea
\left(\begin{array}{c}
\nu_{e} \\
\nu_{\mu} \\
\nu_{\tau}
\end{array}\right)=
U_{PMNS}
\left(\begin{array}{l}
\nu_{1} \\
\nu_{2} \\
\nu_{3}
\end{array}\right),\nnb \,\,\,\,\,\,\,\,\,\,\,\,\,\,\,\,\,\,\,\,\,\,\,\,\,\,\,\,\,\,\,\,\,\,\,\,\,\,\,\,\,\,\,\,\,\,\,\,\,\,\,\,\,\,\,\,\,\,\,\,\,\,\,\,\,\,\,\,\,\,\, (1.1)
\eea
where $U_{PMNS}$ is the neutrino mixing matrix\cite{PMNS}.
%\bea
%U_{PMNS}=U P_{\nu}\nnb %\,\,\,\,\,\,\,\,\,\,\,\,\,\,\,\,\,\,\,\,\,\,\,\,\,\,\,\,\,\,\,\,\,\,\,\,\,\,\,\,\,\,\,\,\,\,\,\,\,\,\,\,\,\,\,\,\,\,\,\,\,\,\,\,\,\,\,\,\,\,\,\,\,\,\,\,\,\,\,\,\,\,\,\,\,\,\,\,\,\,\,\,\,\,\,\,   %(1.2)
%\eea
%with
%\begin{normalsize}
%\bea
%\hskip-0.8cm
%\label{eq:u}
%U=\left(\begin{array}{ccc}
%c_{12} c_{13} & s_{12} c_{13} & s_{13} e^{-\mathrm{i} \delta} \\
%-s_{12} c_{23}-c_{12} s_{13} s_{23} e^{\mathrm{i} \delta} & c_{12} c_{23}-s_{12} s_{13} s_{23} e^{\mathrm{i} \delta} & c_{13} s_{23} \\
%s_{12} s_{23}-c_{12} s_{13} c_{23} e^{\mathrm{i} \delta} & -c_{12} s_{23}-s_{12} s_{13} c_{23} e^{\mathrm{i} \delta} & c_{13} c_{23}
%\end{array}\right)
%,\nnb\eea
%\end{normalsize}
%\bea
%\label{eq:u1}
%^P_{\nu}=diag\left(e^{\mathrm{i} \alpha_{1}},e^{\mathrm{i} \alpha_{2}},1\right)\nnb %\,\,\,\,\,\,\,\,\,\,\,\,\,\,\,\,\,\,\,\,\,\,\,\,\,\,\,\,\,\,\,\,\,\,\,\,\,\,\,\,\,\,\,\,\,\,\,\,\,\,\,\,\,\,\,\,\,\,\,\,\,\,\,\,\,\,\,\,\,\,\,\,\,\,\,\,\,\,   %(1.3)
%\eea
%where $s,c$ denote $\sin$ and $\cos$, respectively. $\theta_{ij}(ij=12,13,23)$ are mixing angles, $\delta$ is  the Dirac CP-violating phase %and $\alpha_{1}, \alpha_{2}$ are Majorana CP phases.
The current experimental data of mixing angles, CP-violation phases and mass square differences for normal and inverted neutrino mass hierarchy have been obtained by means of neutrino oscillation experiments.\footnote {For a review paper,see the Ref.\cite{Esteban} by Esteban ${\it et \,al.}$ where the three flavor mixing angles, one CP-violating phase and two mass differences, which are extracted from a global analysis of current neutrino oscillation data, are listed in Table 3. For more recent paper, see the Ref.\cite{NuFIT}. }

In the flavoured basis, where charged lepton mass matrix is diagonal, Majorana neutrino matrix contains all the information regarding the leptonic CP violation at low energy. At present the measurements of the Dirac CP-violating phase $\delta$ show that
$\delta_{CP} \approx \pi / 2$ or $ \delta_{CP} \approx 3 \pi / 2$
most likely, which means that the leptonic CP violation is maximal most probably. What models of neutrino mass matrix can achieve the maximal CP violation? In this work
we aim to search such models of neutrino mass matrix which can achieve the maximal CP violation and to investigate their features. The Ref.\cite{Singh} gives us a clue to find such models, i.e., to realize ${\rm \Im}(M_{12}M_{23}M_{31})$, where $M=M_\nu M_\nu^\dagger$ , reaches its maximum.  %As shown in the %paper\cite{Singh}, %arXiv.2006.09687, by M.Singh
%$Im(M_{12}M_{23}M_{31}) =J_{CP} \Delta_{sol} \Delta_{atm}^2$
%where $M=M_\nu M_\nu^\dagger$ , so $Im(M_{12}M_{23}M_{31})$ reaches its maximum when $|\sin\delta=1|$.

If the Real part and Imaginary part of a Hermitian matrix $M=M_R+i M_I$ are commutative:

\bea[M_R, M_I]=0,\,\,\,\,\,\,\,\,\,\,\,\,\,\,\,\,\,\,\,\,\,\,\,\,\,\,\,\,\,\,\,\,\,\,\,\,\,\,\,\,\,\,
\,\,\,\,\,\,\,\,\,\,\,\,\,\,\,\,\,\,\,\,\,\,\,\,\,\,\,\,\,\,\,\,\,\,\,\,\,\,\,\,\,\,\,\,\,\,\,\,\,\,\,\,(1.2)\nnb\eea
it is easy to prove that%\footnote{For two Hermitian matrices which can be diagonalized by a same unitary matrix $U$ simultaneously, they %must satisfy an interesting condition\cite{Buras}
%\bea
%M_1M^\dag_2 - M_2M^\dag_1 =0. \,\,\,\, \,\,\,\,\,\,\,\,\,\, \,\,\,\,\,\,\,\,\,\, \,\,\,\,\,\,\,\,\,\, \,\,\,\,\,\,\,\,\,\, \,\,\,\, %\,\,\,\,\,\,\,\,\,\, \,\,\,\,\,\,\,\,\,\,\,\,\,\,\,\,\,\,\,\, \nnb\eea

%Inspired by this, we take the Real component and Imaginary component of a Hermitian matrix $M=M_R+i M_I$ as $M_1$ and $M_2$.}
\bea \label{eq:mrmi} M_R({\rm i}M_I)^\dagger-({\rm i}M_I)M_R^\dagger=0,      \,\,\,\,\,\,\,\,\,\,\,\,\,\,\,\,\,\,\,\,\,\,\,\,\,\,\,\,\,\,\,\,\,\,\,\,\,\,\,\,\,\,\,\,\,\,\,\,\,\,\,\,\,\,\,\,\,\,\,\,\,\,\,\,\,\,\,\,\, \nnb\eea
% Then it is %easy to prove that (1.2) is valid if
%\bea[M_R, M_I]=0.\,\,\,\,\,\,\,\,\,\,\,\,\,\,\,\,\,\,\,\,\,\,\,\,\,\,\,\,\,\,\,\,\,\,\,\,\,\,\,\,\,\,
%\,\,\,\,\,\,\,\,\,\,\,\,\,\,\,\,\,\,\,\,\,\,\,\,\,\,\,\,\,\,\,\,\,\,\,\,\,\,\,\,\,\,\,\,\,\,\,\,\,\,\,\,(1.3)\nnb\eea

which leads to that ${\rm \Im}(M_{12}M_{23}M_{31})$ reaches its maximum.
Therefore, we study the kind of Majorana neutrino mass matrix with the assumption (1.2). Of course, they are phenomenological models. One still needs to have more profound understanding of the flavor structure of neutrinos and flavor symmetry of leptons.

In the bases of the charged lepton mass matrix being diagonal and real, the Majorana neutrino mass matrix $M_{\nu}$ is a complex symmetric mass matrix which contains twelve real parameters. There are two ways to diagonalize the Majorana neutrino mass matrix. One way is directly to diagonalize $M_{\nu}$\cite{Duarah}, the other is to diagonalize the Hermitian matrix $M= M_{\nu} M^\dag_{\nu}$\cite{Aizawa,xy}. In the case of diagonalization of $M_{\nu}$, $U^T_{PMNS}M_{\nu} U_{PMNS}=D_{\nu}$, as pointed out in reference\cite{xy}, there is a parameter mismatching problem since $M_{\nu}$ contains twelve free real parameters but $U_{PMNS}$ and $D_{\nu} = diag(m_1, m_2, m_3)$ totally consist of nine physical parameters. The problem can be solved by choosing a special flavor basis of three neutrino fields to factor out the three unphysical phases of $M_{\nu}$\cite{Aizawa,xy}. In the case of diagonalization of M, $U^+_{PMNS} M U_{PMNS}=D_{\nu}^2 $, since M is a Hermitian matrix, it has advantages of less parameter numbers: 9 free real parameters. We use the special flavor basis of three neutrino fields which makes the three unphysical phases of $M_{\nu}$ be factored out. Using the $M^{\prime}$=$M'_{\nu}M'^+_{\nu}$, where $M'_{\nu}$ is the Majorana neutrino mass matrix in which the three unphysical phases of $M_{\nu}$ have been factored out and then $M'_{\nu}$ has 9 free parameters, then $M^{\prime}$ has only 7 free real parameters since there are two constraint equations for the matrix elements of $M^{\prime}$. For the kind of Hermitian matrix $M^{\prime}$ that we study, there are 4 constraints so that $M^{\prime}$ has 5 free real parameters\footnote{ The $[M_R, M_I] = 0 $ leads there are 4 constraints. It can be proved by a straightforward calculation that the two constraint equations for the matrix elements of $M^{\prime}$ above mentioned are same as the two of the four constraints.}. Therefore, $M'_{\nu}$ has 5 free real parameters. In the paper we hereafter use $M_{\nu}$ and M in stead of $M'_{\nu}$ and $M^{\prime}$ for simplicity.
In the paper, we shall use the latter way, since a Hermitian matrix has advantages of less parameter numbers and two Majorana phases become irrelevant when investigating structure of $M_{\nu}$.
%Furthermore, for the kind of Hermitian matrix $M$ that we study, the condition (1.2) would make M with much fewer parameters.
%which Real part $H_R$ and Imaginary part $H_I$ are commutative,
%it is easy to know a very interesting equation%\footnote{For two Hermitian matrices which can be diagonalized by a same unitary matrix $U$ simultaneously, they must satisfy an interesting condition
%$M_1M^\dag_2 - M_2M^\dag_1 =0$\cite{Buras}.
%Inspired by this, we take the Real component and Imaginary component of a Hermitian matrix as $M_1$ and $M_2$.}

%\bea \label{eq:mrmi} H_R({\rm i} H_I)^\dagger-({\rm i} H_I)H_R^\dagger=0.      %\,\,\,\,\,\,\,\,\,\,\,\,\,\,\,\,\,\,\,\,\,\,\,\,\,\,\,\,\,\,\,\,\,\,\,\,\,\,\,\,\,\,\,\,\,\,\,\,\,\,\,\,\,\,\,\,\,\,\,\,\,\,\,\,\,\,\,\,\, %(1.4)\nnb\eea
%Using Eq.(1.4), one could get Hermitian neutrino mass matrix $M=M_\nu M_\nu^\dagger$ with much fewer parameters further.

In the paper we diagonalize and investigate such kind of Hermitian neutrino mass matrix $M$ of which Real component and Imaginary part are commutative. The virtue of such matrix $M$ is, as pointed out above, with a few parameters, and the mass matrix $M_\nu$ is easily calculated, thus the study of such matrix $M$ should be helpful for investigating flavor mysteries and theoretical models in neutrino physics.
In section 2 we show that a $3\times3$ Hermitian matrix only have five independent real parameters if its Real part and Imaginary part are commutative. The investigation of the kind of Hermitian neutrino mass matrix $M$ whose real and imaginary parts are commutative is carried out in sections 3 and two examples are presented in section 4. Finally, in section 5 we give summary and conclusions. The detailed expressions of the solutions in section 4 are given in Appendix.

\section{Investigation of a kind of Hermitian matrix}
Consider a $3\times3$ Hermitian matrix $H$ of which has the general pattern as
\bea
\label{eq:Hm}
 H =\left(\begin{array}{ccc}  a_1+$i$d_1 &b_1+$i$c_1 & b_2+$i$c_2 \\
b_4+$i$c_4 &a_2+$i$d_2 & b_3+$i$c_3 \\
 b_5+$i$c_5 &b_6+$i$c_6& a_3+$i$d_3 \end{array}\right).\,\,\,\,\,\,\,\,\,\,\,\,\,\,\,\,\,\,\,\,\,\,\,\,\,\,\,\,\,\,\,\,\,\,\,\,\,\,\,\,\,\, (2.1)\nnb    \eea
This Hermitian matrix H contains eighteen real parameters such that
all $a_i, b_j, c_j,d_i (i=1,2,3,j=1,\ldots,6)$ are real.
It is well known that the Hermitian condition leads that
$d_i =0, b_{i+3}= b_i$ and $c_{i+3} =-c_i(i=1, 2, 3)$.
 Thus, Eq.(2.1) now could expressed by two Real matrices of $H_R$ and $H_I$:
\bea
\label{eq:mr}
H&=&H_R+iH_I,\nnb\\
H_R&=&\left(\begin{array}{ccc} a_1 &b_1 & b_2\\
  b_1 & a_2 & b_3 \\
b_2& b_3&a_3 \end{array}\right),\nnb \\ H_I&=&\left(\begin{array}{ccc}
 0 & c_1&c_2\\
-c_1& 0 & c_3\\
-c_2&-c_3 &0
\end{array}\right),\,\,\,\,\,\,\,\,\,\,\,\,\,\,\,\,\,\,\,\,\,\,\,\,\,\,\,\,\,\,\,\,\,\,\,\,\,\,\,\,\,\,\,\,\,\,\,\,\,\,\,\,\,\,\,\,\, (2.2)\nnb \eea where $H_R$ is symmetric, $H_I$ is anti-symmetric. So the number of parameters is reduced from eighteen down to nine.

In this paper we study a kind of Hermitian matrix $H$ of which Real part and Imaginary part are commutative:
\bea [H_R, H_I]=0.\,\,\,\,\,\,\,\,\,\,\,\,\,\,\,\,\,\,\,\,\,\,\,\,\,\,\,\,\,\,\,\,\,\,\,\,\,\,\,\,\,\,\,\,\,\,\,\,\,\,\,\,\,\,\,\,\,\,\,\,\,\,\,\,\,\,\,\,\,\,\,\,\,\,\,\,\,\,\,\,\,\,\,\,\,\,\,\,\,\,\,\,\,\,\,\, (2.3)\nnb\eea

For such a kind of Hermitian matrix, it is straightforward to derive that their Real component and Imaginary component must satisfy a very interesting condition
\bea \label{eq:mrmi} H_R({\rm i}H_I)^\dagger-({\rm i}H_I)H_R^\dagger=0.      \,\,\,\,\,\,\,\,\,\,\,\,\,\,\,\,\,\,\,\,\,\,\,\,\,\,\,\,\,\,\,\,\,\,\,\,\,\,\,\,\,\,\,\,\,\,\,\,\,\,\,\,\,\,\,\,\,\,\,\,\,\,\,\,\,\,\,\,\, (2.4)\nnb\eea

Substituting Eq.(2.2) into Eq.(2.4), we will receive four equations:
\bea \label{eq:four}
b_1&=&c_2c_3(a_1 - a_2)/g,\,\,\,\,\,\,\,\,\,\,\,\,\,\,\,\,\,\,\,\,\,\,\,\,\,\,\,\,\,\,\,\,\,\,\,\,\,\,\,\,\,\,\,\,\,\,\,\,\,\,\,\,\,\,\,\,\,\,\,\,\,\,\,\,\,\,\,\, (2.5) \nnb\\
b_2&=&-c_1c_3(a_1 - a_2)/g, \,\,\,\,\,\,\,\,\,\,\,\,\,\,\,\,\,\,\,\,\,\,\,\,\,\,\,\,\,\,\,\,\,\,\,\,\,\,\,\,\,\,\,\,\,\,\,\,\,\,\,\,\,\,\,\,\,\,\,\,\,\,\,\, (2.6)\nnb\\
b_3&=&c_1c_2(a_1 - a_2)/g,\,\,\,\,\,\,\,\,\,\,\,\,\,\,\,\,\,\,\,\,\,\,\,\,\,\,\,\,\,\,\,\,\,\,\,\,\,\,\,\,\,\,\,\,\,\,\,\,\,\,\,\,\,\,\,\,\,\,\,\,\,\,\,\,\,\,\,\, (2.7)\nnb\\
a_3&=&a_2+(a_1 - a_2)g_1/g,\,\,\,\,\,\,\,\,\,\,\,\,\,\,\,\,\,\,\,\,\,\,\,\,\,\,\,\,\,\,\,\,\,\,\,\,\,\,\,\,\,\,\,\,\,\,\,\,\,\,\,\,\,\,\,\,\,\,\,\,\, (2.8)\nnb
\eea with $g=c_2^2-c^2_3,g_1=c_2^2-c^2_1$
which can be used to further reduce the parameter numbers from nine down to five. We come to the conclusion: a $3\times 3$ Hermitian matrix of which Real component and Imaginary part are commutative has only five real parameters. It is worth noting that for a general Hermitian matrix the conclusion is not valid.
As a matter of convenience we make the following choices: $a_1, a_2, c_i(i=1,2,3)$ are five real parameters, the other four real parameters can be determined from Eqs.(2.5-2.8).

\section{ The general nature of neutrino mass matrix}
% As pointed in \cite{He16}, the particular angles and phase in mixing matrix are parameterization convention dependent. In the paper we %assume the parameterization in \cite{PMNS}, {\it i.e.}, the PMNS matrix which is explicitly given in Eq.(1.3) in Introduction.

The symmetric Majorana neutrino mass matrix $M_\nu$ can be parameterized by
         \bea  M_{\nu}= \left(\begin{array}{ccc}
a&b&c\\
b&d&e\\
c&e&f\end{array}\right)\,\,\,\,\,\,\,\,\,\,\,\,\,\,\,\,\,\,\,\,\,\,\,\,\,\,\,\,\,\,\,\,\,\,\,\,\,\,\,\,   \,\,\,\,\,\,\,\,\,\,\,\,\,\,\,\,\,\,\,\,\,\,\,\,\,\,\,\,\,\,\,\,\,\,\,\,\,\,\,\, (3.1) \nnb \eea
which has nine real parameters. Although $M_\nu$ can be diagonalized by
$U_{\rm PMNS}$\cite{Duarah}, a simpler method is to diagonalize the Hermitian matrix $M=M_\nu M_\nu^\dagger $ \cite{Aizawa},
        \bea M= \left(\begin{array}{ccc}
A&B&C\\
B^*&D&E\\
C^*&E^*&F\end{array}\right)   \,\,\,\,\,\,\,\,\,\,\, \,\,\,\,\,\,\,\,\,\,\,\,\,\,\,\,\,\,\,\,\,\,\,\,\,\,\,\,\,\, \,\,\,\,\,\,\,\,\,\,\,\,\,\,\,\,\,\,\,\,\,\,\,\,\,\,\,\,\,\,\,\,\,\, (3.2) \nnb \eea
with \bea
A&=&|a|^2+|b|^2+|c|^2, \,\,\,\,B=a b^*+b d^*+c e^*,\,\,\,\,\nnb\\ C&=&a c^*+b e^*+c f^*,\,\,\,\,
D=|b|^2+|d|^2+|e|^2, \,\,\,\, \nnb\\ E&=&b c^*+d e^*+e f^*,\,\,\,\, F=|c|^2+|e|^2+|f|^2.  \,\,\,\,\,\,\,\,\,\,\,\,\,\,\,\, (3.3)\nnb \eea
As it is obvious, the Hermitian matrix $M$ has nine real parameters.
In the paper we investigate a kind of Hermitian matrix $M=M_\nu M_\nu^\dagger$ of which Real component and Imaginary part are commutative so that we can use the %method and
 results obtained in last section.
We perform diagonalization by:
   \bea U^\dagger_{\rm PMNS}M U_{\rm PMNS}=  \left(\begin{array}{ccc}
m_1^2&0&0\\
0&m_2^2&0\\
0&0&m_3^2\end{array}\right),  \,\,\,\,\,\, \,\,\,\,\,\,\,\,\,\,\,\,\,\,\,\,\,\,\,\,\,\,\,\,\,\,\,\,\,\,\,\,(3.4) \nnb\eea
where $m_1,m_2,m_3$ are real.
%For the sake of convenience, we assume the Majorana CP phases are zeros.
From Eq.(3.2) and Eq.(3.4), one could get these relations:
\bea
m_1^2&=&\xi_1-\chi t_{12} ,  \,\,\,\, \,\,\,\,\,\,\,\,\,\,\,\,\,\,\,\,\,\,\,\,\,\,\, \,\,\,\,\,\,\,\,\,\,\,\,\,\,\,\,\,\,\  \,\, \,\,\,\,\,\,\,\,\,\,\,\,\,\,\,\,\,\,\,\,\,\,\,\,\,\,\,\,\,\,\,\,\,\,\,(3.5)  \nnb\\
m_{2}^2&=&\xi_{2}+\chi t_{12} =\xi_1+\frac{\chi}{t_{12}},   \,\,\,\,\,\,\,\,\,\,\,\,\,\,\,\,\,\,\,\,\,\,\,\,\,\,\,\,\,\,\,\,\,\, \,\,\, \,\,\,\,\,\,\,\,\,\,\,\,\,\,\,\,\,\,\,\,(3.6)  \nnb \\
m_{3}^2&=&c^2_{13}(\xi_3+t_{13}^2 A+t_{13}|y|2\cos \beta), \,\,\,\, \,\, \,\,\,\,\,\,\,\,\,\,\,\,\,\,\,\,\,\,\,\,\,\,\,\,\,\,\,\,\,\, \,\,(3.7)  \nnb \\
\xi_1&= &A-|y|\cos\beta t_{13} ,\,\,\,\, \,\,\,\,\,\,\,\,\,\,\,\,\,\,\,\,\,\,\,\,\,\,\, \,\,\,\,\,\,\,\,\,\,\,\,\,\,\,\,\,\,\,\,\,\,\, \,\,\,\,\,\,\,\,\,\,\,   \,\,\,\,\,\,\,\,\,(3.8)  \nnb \\
 \chi&=&\frac{c_{23}{\rm \Re}(B)-s_{23}{\rm \Re}(C)}{c_{13}} , \,\,\,\,\,\,\,\, \,\,\,\,\,\,\,\,\,\,\,\,\,\,\,\,\,\,\,\,\,\,\, \,\,\,\,\,\,\,\,\,\,\,\,\,\,\,\,\,\,\,\,\,\,\,\,\,\,\,\,  \,\, (3.9)  \nnb\\
\xi_2&=&D+{\rm \Re}(E)(-t_{23}),  \,\,\,\,\,\,\,\, \,\,\,\,\,\,\,\,\,\,\,\,\,\,\,\,\,\,\,\,\,\,\, \,\,\,\,\,\,\,\,\,\,\,\,\,\,\, \,\,\,\,\,\,\,\,\,\,\,\,\,\,\,\,\,\,\,\,(3.10)  \nnb\\ \xi_3&=&A+\frac{2|y|(\cos (2\theta_{13})\cos\beta+{\rm i}\sin\beta)}{\sin(2\theta_{13})} ,\,\,\,\,\,\,\,\,\,\,\,\,\,\,\, \,\,\,\,\,\,\,(3.11)  \nnb\eea  \\
where \\ \bea
y&=&s_{23}B +c_{23} C=|y|e^{\mathrm{i}\alpha},\,\,\,\,  \beta=\alpha + \delta, \,\,\,\,\,\,\,\,\,\,\,\,\,\,\,\,\,\,\,\,\,\,\,\,\,\,\,\,\,\,\,\,\,\,\,\,\,\,\,\, (3.12)  \nnb \\
\xi_2+\xi_3&=&D+F,\,\,\,\,\,\,\,\,\,\,\,\,\,\,\,\,\,\,\,\,\,\,\,\,\,\,\,\,\,\,\,\,\,\,\,\,\,\,\,\,\,\,\,\,\,\,\,\,\,(3.13) \nnb \eea \\
and \bea t_{ij}=\tan\theta_{ij}=\frac{\sin\theta_{ij}}{\cos\theta_{ij}},\,\, s_{ij}=\sin\theta_{ij},\,\, c_{ij}=\cos\theta_{ij}, \,\,\,\,\, %t=tan=\frac{\sin}{\cos}.\,\,\,\,\, \,\,
\,\,\,\,\,\,\,\,\,\,\,\,\,\,\,\,\,\,\,\,\nnb\eea \\
with $\theta_{ij}$  (i j = 12, 13, 23) are mixing angles. \\
And one also has\bea
{\rm \Im}(B)&=&t_{23} {\rm \Im}(C) , \,\,\,\,\,\,\,\,\,\,\,\,\,\,\,\, \,\,\,\,\,\,\,\,\,\,\,\,\,\,\,\,\,\,\,\,\, \,\,\,\,\,\,\,\,\,\, \,\,\,\,\,\,\,\,\,\,\,\,\,\,\,\,\,\,\,\,\,\,\,\,\,\,\,\,\,\,\,\,\,   (3.14) \nnb \\{\rm \Im}(E)&=&s_{13} \chi \sin \delta ,\,\,\,\,\,\,\,\,\,\,\,\,\,\,\,\,\,\, \,\, \,\,\,\,\,\,\,\,\,\,\,\,\,\,\,\,\,\,\,\,\,\,\,\,\,\,\, \,\,\,\,\,\,\,\,     \,\,\,\,\,\,\,\,\,\,\,\,\,\,\,\,\,\,\,\, (3.15)\nnb\\
\cos \delta&=&\frac{\sin (2\theta_{23})\frac{F-D}{2}-\cos (2\theta_{23}){\rm \Re}(E)}{ s_{13} \chi }.  \,\,\,\,\,\,\,\,\,\,\,\,\,\,\,\,\,\,\,\,\,\,\,\,\,\,(3.16)\nnb\eea

It is important to note that Eq.(3.7) means that $\xi_3$ must be real. Then, from Eq.(3.11), we have
the result of \bea
 \sin \beta=0, \beta=\pm(n\pi), n=0,1,2....\,\,\,\,\,\,\,\,\,\,\, \,\,\,\,\,\,\,\,\,\,\,\,\,\,\,\,\,\,\,\,\,\,\,\,\,\,\,\, \,\,\,\,\,\, (3.17)\nnb\eea
Furthermore, $\xi_1,\xi_2$ satisfies the constraint \bea
(\xi_2-\xi_1)tg(2 \theta_{12})=2\chi. \,\,\,\,\,\,\,\,\,\,\,\,\,\,\,\,\,\,\,\,\,\,\,\,\, \,\,\,\,\,\,\,\,\,\,\,\,\,\,\,\,\,\,\,\,\,\,\,\,\,\,\,\,\,\,\,\, \,\,\,\, \,\,\,\,\,\,\,\,\,\,\, (3.18)\nnb\eea

 According to the analysis of the previous section, %for a $3\times 3$ Hermitian matrix, if its Real component and Imaginary part are %commutative, with imposing four conditions of Eqs.(2.5-2.8), we could express this matrix form in terms of only five real parameters.
a $3\times 3$ Hermitian matrix of which Real component and Imaginary part are commutative has only five real parameters. That is, there are four Eqs.(2.5-2.8) which mean there are the four constraints as follows

 \bea
{\rm \Re}(B) {\rm \Im}(B)&=&-{\rm \Re}(C) {\rm \Im}(C) = {\rm \Re}(E) {\rm \Im}(E), \,\,\,\,\,\,\, \,\,\,\,\,\,\,\,\,\,\,\,\,\,\,\,\,\,\,\,\,\,\,\,\,  (3.19)\nnb \\
A-D&=&\frac{ {\rm \Re}(E) {\rm \Im}(C)+ {\rm \Re}(C) {\rm \Im}(E) }{{\rm \Im}(B)} ,\,\,\,\,\,\,\, \,\,\,\,\,\,\,\,\,\,\,\,\,\,\,\,\,\,\,\,\,\,\,\,\,\,\,\,\,          (3.20)\nnb\\
F-D&=&\frac{{\rm \Re}(C) {\rm \Im}(B)+ {\rm \Re}(B) {\rm \Im}(C) }{{\rm \Im}(E)}.  \,\,\,\,\,\,\,\,\,\,\,\,\,\,\,\,\,\,\,\,\,\,\,\,\,\,\,\,\,\,\,\,\,\,\,\,\,\,           (3.21)\nnb
\eea
We can easily obtain the five independent real parameters by using the above four constraints (so $M_\nu$ has 8 real parameters) and transform them into two parameters plus three mixing angles. And we obtain two important conclusions when the four constraints are imposed.

From Eq.(3.14) and Eq.(3.19), it is straightforward to get the relation of \bea{\rm \Re}(C)=-t_{23} {\rm \Re}(B). \,\,\,\,\,\,\, \,\,\,\,\,\,\, \,\,\,\,\,\,\,\,\,\,\, \,\,\,\,\,\,\,\,\,\,\, \,\,\,\,\,\,\,\,\,\,\, \,\,\,\,\,\,\,\,\,\,\,\,\,\,\,\,\,\,\,\,\,\,\, \,\,\,\,\,\,\,\,\,\,\,\,\,\,\,\, \, (3.22) \nnb\eea
Considering Eqs.(3.12) and (3.22), one can easily derive that the Real component of $y$ is zero:
 \bea {\rm \Re} (y) =0,\,\,\,\,\,\,\,\, \,\,\,\,\,\,\,\,\,\,\, \,\,\,\,\,\,\,\,\,\,\,\,\,\,\,\,\, \,\,\,\,\,\,\,\,\,\,\, \,(3.23') \nnb\eea
  {\it {\it i.e.}},\bea
\cos \alpha=0, \alpha=\pm(\frac{\pi}{2}+m\pi), \,\,\,\,m=0,1,2.... \,\,\,\,\,\, \,\,\,\,\,\,\,\,\,\,\, \,\,\,\,\,\,\,\,\,\,\,\,\,(3.23) \nnb\eea
provided that the modulus of y is not equal to zero,
and  \bea {\rm \Im} (y) =\frac{{\rm \Im}(C)}{c_{23}}.\,\,\,\,\,\,\,\,\, \,\,\,\,\,\,\,\,\,\,\, \,\,\,\,\,\,\,\,\,\,\, \,\,\,\, \,\,\,\,\,\,\, \,\,\,\,\, \,\,\,\,\,\, \,\,\,\,\, \,\, \,\,\,\,\,\,\,\,\,\,\,\,\,\,\,\,\,\,\,\,\,\,\,\,\,\,\,\,\,\,\,\,\,\,\,\,\,\,\, (3.24)\nnb\eea

Eqs.(3.12),(3.17) and Eq.(3.23) lead to
\bea\cos\delta=0, \delta =\pm(\frac{\pi}{2}+p\pi), \,\,\,\,p=0,1,2....  \,\,\,\,\,\,\,\,\,\,\,\,\,\,\,\,\,\,\,\,\,\,\,\,\,\,\,\,\,\,\,\,\,  (3.25)\nnb\eea
Therefore, due to Eq.(3.16), we have
\bea\frac{\sin (2\theta_{23})\frac{F-D}{2}-\cos (2\theta_{23}){\rm \Re} (E)}{ s_{13} \chi }=0. \,\,\,\, \,\,\,\,\,\,\,\,\,\,\,\,\,\,\,\,\,\,\,\,\,\,\,\,\,\,\,\,\,\,\,\,\,\,\,\,\,\,\,\,\,   (3.26)\nnb\eea

Furthermore, The fourth equation, Eq.(3.21), of the constraints, can be simplified as \bea F-D=\frac{\cos (2\theta_{23}){\rm \Im}(C)}{c_{23}t_{13}\sin\delta}. \,\,\,\,\,\,\,\,\,\,\,\,\,\,\,\,\, \,\,\,\,\,\,\,\,\,\, \,\,\,\,\,\,\,\,\,\, \,\,\, \,\,\,\,\,\,\,\,\,\,\,\,\,\,\,\,\,\,\,\,\,\,\,\,\,\,\,\,\,\,   (3.27)\nnb\eea
Substituting Eq.(3.27) into Eq.(3.26), one has
\bea \cos (2\theta_{23})(\frac{\sin (2\theta_{23}){\rm \Im}(C)}{2c_{23}t_{13}\sin\delta}-{\rm \Re} (E))=0,     \,\,  \,\,\,\,\,\,\,\,\,\,\,\,\,\,\,\,\,\,\,\,\,\,\,\,\,\,\,\,\,\,\,\,\,\,\,\,\,    (3.28)\nnb\eea
since $s_{13}\chi$ is not equal to zero.
To satisfy Eq.(3.28), there are the following cases:\bea  \,\,\,\,\,\,\,\,\,\,\,\,\,\,\,\,a) \,\,\,\cos (2\theta_{23})=0,   \,\,\,\,\,\,\,\,\,\,\,\,\,\,\,\,\,\,\,\,\,\,\,\, \,\,\,\,\,\,\,\,\,\,\,\, \,\,\,\,\,\,\,\,\,\,\,\,\,\,\,\,\,\,\,\,\,\,\,\,\,\,\,\,\,(3.29a)  \nnb\eea or \bea\,\,\,\,\,\,\,\,\,\,\,\,  b)\,\,\,
{\rm \Re} (E)=\frac{\sin (2\theta_{23}){\rm \Im }(C)}{2c_{23}t_{13}\sin\delta},\,\,\,\,\,\,\,\,\,\,\,\, \,\,\,\,\,\,\, \,\,\,\,\,\,\,\,\,\,\,\,\,\,\,\,\,\,\,\,\,\,\,\,\,\,\,\,\,\,\,\,\,  (3.29b) \nnb\eea or both factors are zeros.

We choose $A$, ${\rm \Im}(C)$, and three mixing angles as five real parameters for specific. In terms of them, other real parameters could expressed as:
  \bea{\rm \Re} (E) &=& \frac{{\rm \Im}(C) s_{23}}{t_{13}\sin\delta}, \,\,\,\,\,\,\,\,\,\,\,\,\,\,\,\,\,\,\,\,\,\,\,\,\,\,\,\,\,\,\,\,\,\,\,\,\,\,\,\,\,\,\,\,\,\,\,\,\,\,\,\,\,\,\,\,\,\,\,\,\,\,\,\,\,\,\,\,\,\,\,\,  (3.30)\nnb \\
  {\rm \Re}(B)&=& \frac{{\rm \Im}(E) c_{23}}{ t_{13} \sin \delta},    \,\,\,\,\,\,\,\,\,\,\,\,\,\,\,\,\,\,\,\,\,\,\,\,\,\,\,\,\, \,\,\,\,\,\,\,\,\,\,\,\,\,\,\,\,\,\,\,\,\,\,\,\,\,\,\,\,\,\,\,\,\,\,\,\,\,\,\,\,\,\,\,\,      (3.31)\nnb \\
{\rm \Re}(C)&=&\frac{-{\rm \Im}(E) s_{23}}{ t_{13} \sin \delta}, \,\,\,\,\,\,\,\,\,\,\,\,\,\,\,\,\,\,\,\,\,\,\,\,\,\,\,\,\,\,\,\,\,\,\,\,\,\,\,\,\,\,\,\,\,\,\,\,\,\,\,\,\,\,\,\,\,\,\,\,\,\,\,\,\,\,\,\,\,\,   (3.32)\nnb\\
 A-D&=&\frac{-2|y|\cos (2\theta_{13}) \cos\beta}{\sin (2\theta_{13})} +\frac{{\rm \Re}(E)}{t_{23}}. \,\,\,\,\,\,\,\,\,\,\,\,\,\,\,\,\,\,\,\,\,\,\,\,\,\, (3.33) \nnb\eea
Besides, combining Eqs.(3.19),(3.20) and Eq.(3.32), one could get the following relation by introducing the variable $g={\rm \Im}(C)^2-{\rm \Im}(E)^2$: \bea g=(A-D)\frac{{\rm \Im}(C) (t_{13}\sin\delta)}{c_{23}}. \,\,\,\,\,\,\,\,\,\,\,\,\,\,\,\,\,\,\,\,\,\,\,\,\,\,\,\,\,\,\,\,\,\,\,\,\,\,\,\,\,\,\,\,\,\,\,\,\,\,\,\,\,\,\,\,\,\,\,\,\,   (3.34)  \nnb\eea

Considering Eqs.(3.33) and (3.34), one obtains the equation:\bea {\rm \Im}(E)^2=|y|^2(2|c_{23}|^2-1+t_{13}^2).  \,\,\,\,\,\,\,\,\,\,\,\,\,\,\,\,\,\,\,\,\,\,\,\,\,\,\,\,\,\,\,\,\,\,\,\,\,\,\,\,\,\,\,\,\,\,\,\,\,\,\,\,\,\,\,
 (3.35)\nnb\eea
Thus, the collection of Eqs.(3.14,3.27,3.30-33,3.35) explicitly gives the Hermitian Majorana neutrino mass matrix $M$ in terms of $A$, ${\rm \Im}(C)$, and three mixing angles.

%Before we close this section, some comments are in place.
Moreover, we would like to determine the parameters $A$ and ${\rm \Im}(C)$ by using mass squared differences.
From Eqs.(3.5-3.7), we have\bea \Delta m_{21}^2 &=&|m_2^2-m_1^2| =|\frac{\chi(1+t_{12}^2)}{t_{12}}| , \,\,\,\,\,\,\,\,\,\,\,\,\,\,\,\,\,\,\,\,\,\,\,\,\,\,\,\,\,\,\,\,\,\,\,\,\,\, (3.36)\nnb\\
 \Delta m_{31}^2 &=&|m_3^2-m_1^2 |=|\frac{|y|2\cos\beta}{\sin (2\theta_{13})}+\chi t_{12}|.\,\,\,\,\,\,\,\,\,\,\,\,\,\,\,\,\,\,\,\,\,\,\,\,\,  (3.37)\nnb\eea
Then we can use the measured values of mass squared differences to get the parameters
${\rm \Im} (C)$ and $\chi$. However, Eqs.(3.8),(3.10),(3.18) lead to that $\chi$ depends $(A-D)$ %But the two mass squared differences, %Eqs.(4.1),(4.2), can determine $\chi$  as well as ${\rm \Im} C$ only
so that one free parameter A or D still not fixed.
 Thus, there is only one parameter A (or D) in the Hermitian neutrino mass matrix M.

 In summary of this section, we emphasize that $\delta = \pm\frac{\pi}{2}$ of Eq.(3.25) implies the maximal strength of CP violation in neutrino oscillations for given values of $\theta_{12}$, $\theta_{13}$ and $\theta_{23}$ ({\it i.e.}, the leptonic Jarlskog parameter is maximal in this case\cite{Jarlskog}), which is a robust result for the kind of M that we study in the paper. The constraint, the first equation of (3.19), and the constraint, Eq.(3.14), due to diagonalization lead to $\cos \delta$ = 0. That is, the condition to have maximum CP violation is explicitly realized in this kind of models. One of possibilities to make Eq.(3.38) valid is $\cos (2\theta_{23})=0$. And in this case,
Eq.(3.27) leads to the relation of F = D. % i.e., Eq.(3.29a), \footnote{Because $\cos (2\theta_{23})=0$ is not a necessary condition for $\cos \delta=0$, we call it a prediction, not a result.} is just one of probabilities, {\it {\it i.e.}}, both factors of Eq.(3.28) are equal zero, since Eq.(3.29b) has been satisfied (see Eq.(3.30)). And in this case, Eq.(3.27) leads to the relation of $F=D$.

\section{More discussions on Majorana neutrino mass matrix}
%We now make some comments about the general Majorana neutrino mass matrix M¦Í (with 12 parameters) based only on the assumption considered %in eq.(2.3). Indeed, $M_\nu=U D_\nu U^T$ , $D_\nu =dig{m_1,m_2,m_3}$. U is the PMNS matrix.

Eqs.(3.5-3.7) can be reduced to

\bea
m_{1}^{2}&=&A-|y| \cos \beta t_{13}-\chi t_{12},\,\,\,\,\,\,\,\,\,\,\,\,\,\,\,\,\,\,
\,\,\,\,\,\,\,\,\,\,\,\,\,\,\,\,\,\,\,\,\,\,
\,\,\,\,\,\,\,\,\,\,\,\,(3.38)\nnb\\
m_{2}^{2}&=&%&\xi_{1}+\chi / t_{12}=
A-|y| \cos \beta t_{13}+\chi / t_{12},\,\,\,\,\,\,\,\,\,\,\,\,\,\,\,\,\,\,\,\,\,\,\,
\,\,\,\,\,\,\,\,\,\,\,\,\,\,\,\,\,
\,\,\,\,\,\,\,\,\,(3.39)\nnb\\
m_{3}^{2}&=&A+|y| \cos \beta / t_{13},\,\,\,\,\,\,\,\,\,\,\,\,\,\,\,\,\,\,\,\,\,\,\,
\,\,\,\,\,\,\,\,\,\,\,\,\,\,\,\,\,\,\,\,\,\,\,\,\,\,
\,\,\,\,\,\,\,\,\,\,\,\,\,\,\,\,\,\,\,(3.40)\nnb\eea

%$\mathbf{m}_{1}{ }^{2}=\mathbf{A}-|\mathbf{y}| \cos \beta \mathbf{t}_{13}-\mathbf{X}\chi \mathbf{t}_{12},\,\,\,\,\,\,(3.38)\nnb\\$
%$\mathbf{m}_{2}{ }^{2}=\xi_{1}+X / \mathrm{t}_{12}=\mathbf{A}-|\mathrm{y}| \cos \beta \mathrm{t}_{13}+\mathrm{X}\chi / %\mathrm{t}_{12},\,\,\,\,(3.39)\nnb\\$
%$\mathrm{~m}_{3}{ }^{2}=\mathrm{A}+|\mathrm{y}| \cos \beta / \mathrm{t}_{13},\,\,\,\,\,\,(3.40)\nnb\\$

After inputting experimental data of three mixing angles and two mass squared differences, all three masses, $m_1,m_2,m_3$, depend on the free parameter A (or D) which is only single free parameter in our case, as obviously seen from Eqs.(3.38-3.40). Because $M_\nu=U D_\nu U^T$, $D_\nu =diag(m_1,m_2,m_3)$, U is the $PMNS$ matrix, we have that the general Majorana neutrino mass matrix $M_\nu$ depends on the free parameter A (or D) only. That is, for the  kind of Hermitian matrix $M=M_\nu M_\nu^\dagger$ of which Real component and Imaginary part are commutative, the general Majorana neutrino mass matrix $M_\nu$ is known except for a unknown parameter A after inputting the experimental data from neutrino oscillations. Such, one can not determine neutrino mass scale unless to input one more experimental data. For example, from experiments on the search for the neutrinoless double $\beta$
decay\cite{gan}, one has
%0.08 eV \leq|m_{ \beta \beta}| \leq 0.81 eV  \nnb
$m_{\beta \beta} \leq 0.165 eV $,
%$0.08¡Ü|m_{\beta \beta}|¡Ü0.81$ (see Table II in the paper, arXiv.2008.02110) which leads to m1=, and then A=
which leads to $m_{1} \leq 0.15 eV $
%0.079 eV \leq m_{1} \leq 2.24 eV  \nnb
and then $A \leq 0.023 eV^2$.
%6.42\times 10^{-3} eV^2 \leq A \leq 5.02 eV^2  \nnb

Based on the Eqs.(3.38-3.40), it is easy to get \\

1)$A > 0$;\\

2)the constraints to model parameters from the neutrino mass hierarchy: for NMO (normal
mass ordering), $\cos \beta=+1,\chi>0$; for IMO (inverted mass ordering), $\cos \beta=-1, \chi<0$.\\

Moreover, Eqs.(3.18), (3.33) lead to the sum rule of mixing angles \bea
\frac{2\sqrt{2|c_{23}|^2-1+t_{13}^2}}{c_{13}tg(2\theta_{12})}
+\epsilon=1, \,\,\,\,\,\,\,\,\,\,\,\,\,\,\,\,\,\,\,\,\,\,\,\,\,\,\,\,\,\,\,\,\,\,\,\,\,\,\,\,\,\,\,\,\,\,\,\,\,\,\,\,\,\,\,\,\,  (3.41)\nnb\eea
where $\epsilon=\pm 1$. Using data of the current global fits\cite{Esteban,NuFIT}
, it follows that the sum rule is in agreement with data at 3 sigma level for both mass hierarchies. For the sake of satisfying Eq.(3.41), one has $\epsilon=+1$ so that $\cos(2\theta_{23})<0$, therefore,
\bea
  \pi/4 < \theta_{23} < 3\pi/4,
  \,\,\,\,\,\,\,\,\,\,\,\,\,\,\,\,\,\,\,\,\,\,\,\,\,\,\,\,\,\,\,\,\,\,\,\,\,\,\,\,\,\,\,\,\,\,\,\,\,\,\,\,\,\,\,\,\,\,\,\,\,\,\,\,\,\,\,(3.42a)\nnb
\eea
or
\bea
-3\pi/4<\theta_{23} < - \pi/4. \,\,\,\,\,\,\,\,\,\,\,\,\,\,\,\,\,\,\,\,\,\,\,\,\,\,\,\,\,\,\,\,\,\,\,\,\,\,\,\,\,\,\,\,\,\,\,\,\,\,\,\,\,\,\,\,\,\,\,\,\,\,\,\,(3.42b) \nnb
\eea
It is expected that the prediction, Eq.(3.42), will be verified in the future experiments.

\section{ Some examples }

%From Eqs.(3.5-3.7), we have\bea \Delta m_{21}^2 &=&|m_2^2-m_1^2| =|\frac{\chi(1+t_{12}^2)}{t_{12}}| , %\,\,\,\,\,\,\,\,\,\,\,\,\,\,\,\,\,\,\,\,\,\,\,\,\,\,\,\,\,\,\,\,\,\,\,\,\,\,\,\, (4.1)\nnb\\
 %\Delta m_{31}^2 &=&|m_3^2-m_1^2 |=|\frac{|y|2\cos\beta}{\sin (2\theta_{13})}+\chi %t_{12}|.\,\,\,\,\,\,\,\,\,\,\,\,\,\,\,\,\,\,\,\,\,\,\,\,\,\,\,  (4.2)\nnb\eea
%Then we can use the measured values of mass differences to get the parameters
%${\rm \Im} C$ and $\chi$. And Eqs.(3.8),(3.10),(3.18) lead to that $\chi$ depends $(A-D)$. But the two mass squared differences, %Eqs.(4.1),(4.2), can determine $\chi$  as well as ${\rm \Im} C$ only
%so that one free parameter D or A still not fixed.
% Thus, there is only one parameter D (or A) in the neutrino mass matrix M.

It is straightforward to obtain elements of the matrix $M_{\nu}$, which are useful to compare with theoretical models and give classification and some constraints on them, in terms of above five real parameters by solving Eq.(3.3). For the purpose of seeing what type of matrix $M_{\nu}$ belongs to the kind of matrix which is studied in this paper, we study some specific examples.

1)Model I

  An extended version of $\mu-\tau$ symmetric mass matrix which accommodate non-zero $\theta_{13}$ along with maximal CP violation is given in \cite{Duarah}
\bea M_\nu&=&
\left(\begin{array}{ccc}  a & b & -b^* \\
b&d &e \\
-b^*&e &d^* \end{array}\right),\,\,\,\,\,\,\,\,\,\,\,\,\,\,\,\,\,\,\,\,\,\,\,\,\,\,\,\,\,\,\,\,\,\,\,\,\,\,\,\,\,\,\,\,\,\,\,\,\,\,\,\,\,\,\,\, (4.1)\nnb
\eea here $a$ and $e$ are real. The neutrino mass matrix has been discussed in a number of works with discrete flavour symmetry models \cite{king,Scott,Babu,Ma,Grimus,he,Fukuyama,xing,Popov}. If the matrix $M=M_\nu M_\nu^\dagger $ belongs the
kind of matrix of which Real component and Imaginary part are commutative, we can get the expression of the matrix $M_\nu$ in terms of one free parameter and experimental data on neutrinos.

Applying Eqs.(3.3) and (4.1), one has the following equations:\bea
{\rm \Re}(B)&=&-{\rm \Re}(C),\,\,\,\,\,\,\,\,\,\,\,\,\,\,\,\,\,\,\,\,\,\,\,\,\,\,\,\,\,\,\,\,\,\,\,\,\,\,\,\,\,\,\,\,\,\,\,\,\,\,\,\,\,\,\,\,\,\,\,\,\,\,\,\,\,\,\,\,\,\,\,\,\,\,\,\,\,\,\,\,\,\,\, (4.2a)\nnb\\
 {\rm \Im}(B)&=&{\rm \Im}(C),\,\,\,\,\,\,\,\,\,\,\,\,\,\,\,\,\,\,\,\,\,\,\,\,\,\,\,\,\,\,\,\,\,\,\,\,\,\,\,\,\,\,\,\,\,\,\,\,\,\,\,\,\,\,\,\,\,\,\,\,\,\,\,\,\,\,\,\,\,\,\,\,\,\,\,\,\,\,\,\,\,\,\,\,\,\, (4.2b)\nnb\\
   F&=&D.\,\,\,\,\,\,\,\,\,\,\,\,\,\,\,\,\,\,\,\,\,\,\,\,\,\,\,\,\,\,\,\,\,\,\,\,\,\,\,\,\,\,\,\,\,\,\,\,\,\,\,\,\,\,\,\,\,\,\,\,\,\,\,\,\,\,\,\,\,\,\,\,\,\,\,\,\,\,\,\,\,\,\,\,\,\,\,\,\,\,\,\,\,\,\,\, (4.3)\nnb\eea
  Accordingly, using the method described in section 3, it follows that due to Eq.(3.23), one could receive
   $\cos\delta=0$ and due to Eq.(3.14),
  one has
   $t_{23} =1$ which means $cos(2\theta_{23})=0$, as expected.
   Thus Eq.(3.3) can be reduced to the following six equations.\\
\bea
&&-x_1^2+x_2^2+2y_1z_2=a_1 , \,\,\,\,\,\,\,\,\,\,\,\,\,\,\,\,\,\,\,\,\,\,\,\,\,\,\,\,\,\,\,\,\,\,\,\,\,\,\, \,\,\, \,\,\,  \,\,\,\, \,\,\, \,\,\, \,\,\, \,\,\, (4.4)\nnb\\
  &&2x_1x_2-2y_2z_2=a_2,\,\,\, \,\,\,  \,\,\,\,\,\,\,\,\,\,\,\,\,\,\,\,\,\,\,\,\,\,\,\,\,\,\,\,\,\,\,\,\,\,\,\,\,\,\, \,\, \,\,\, \,\,\, \,\,\, \,\,\, \,\,\, \,\,\, \,\,\,  (4.5) \nnb \\
  &&(z_1-z_2)x_1+x_1y_1+x_2y_2=b_1,  \,\,\,\,\,\,\,\,\,\,\,\,\,\,\,\,\,\,\,\,\,\,\,\,\,\,\,\,\,\,\,\,\,\,\,\,\,\,\, \,\,\,(4.6)\nnb \\
&&(z_1-z_2)x_2+x_1y_2-x_2y_1=b_2,  \,\,\, \,\,\,\,\,\,\,\,\,\,\,\,\,\,\,\,\,\,\,\,\,\,\,\,\,\,\,\,\,\,\,\,\,\,\,\,\,\,\, (4.7) \nnb \\  &&z_1^2+2(x_1^2+x_2^2)=c_1,\,\,\, \,\,\,  \,\,\, \,\,\, \,\,\, \,\,\,\,\,\,\,\,\,\, \,\,\, \,\,\, \,\,\, \,\,\,\,\,\,\,\,\,\,\,\,\,\,\,\,\,\,\,\,\,\,\,\,\,\,\,\,\,\,\, \, (4.8)\nnb\\&&
x_1^2+x_2^2+y_1^2+y_2^2+z_2^2=c_2,  \,\,\,\,\,\,\,\,\,\,\,\,\,\,\,\,\,\,\,\,\,\,\,\,\,\,\,\,\,\,\,\,\,\,\,\,\,\,\,\,\, \,\,\,\,\,\,\,      (4.9)\nnb
\eea
where $x_1={\rm \Re}(b), x_2={\rm \Im}(b), y_1={\rm \Re}(d), y_2={\rm \Im}(d), z_1=a, z_2=e$ and $a_1={\rm \Re}(E), a_2={\rm \Im}(E), b_1={\rm \Re}(B), b_2={\rm \Im}(B),c_1=A, c_2=D$. From Eqs.(3.30),(3.35), we have the relation of $a_1$ and $a_2$:
$a_1=\pm w^2a_2$ with $w=1/(t_{13}\sqrt{2})$. And Eqs.(3.31),(3.35) lead to the relation of $b_1$ and $b_2$: $b_1=\pm b_2.$  So, the equations (4.4-4.9) can be simplified as:
\bea \label{eq}
\left\{\begin{array}{c}x^2_1-x^2_2=2y_1z_2\mp w b_2, \,\,\,\,\,\,\,\,\,\,\,\,\,\,\,\,\,\,\,\,\,\,\,\,\,\,\,\,\,\,\, \,\,\,\,\,\,\,\,\,\,\,\,\,\,\,\,\,\,\,\,\,\,\,\,\,\,\,\,\,\,\,\,(4.10)\\
2x_1x_2-2y_2z_2=b_2/(\pm w), \,\,\,\,\,\,\,\,\,\,\,\,\,\,\,\,\,\,\,\,\,\,\,\,\,\,\,\,\,\,\,\,\,\,\,\,\,\,\,\,\,\,\,\,\,\,\,\,\,\,\,\,\,\,\,\,\,
(4.11)\\ \pm x_2-x_1=\pm y_2 w+y_1/(\pm w), \,\,\,\,\,\,\,\,\,\,\,\,\,\,\,\,\,\,\,\,\,\,\,\,\,\,\,
\,\,\,\,\,\,\,\,\,\,\,\,\,\,\,\,\,\,\,\,\,\,\,\,(4.12)\\
(\pm b_2-b_2)=(z_1-z_2-y_2)(x_1-x_2)+y_1(x_1+x_2),
\,\,\,(4.13) \\
z^2_1+2(x_1^2+x_2^2)=c_2+b_2(w^2-1)/(\pm w), \,\,\,\,\,\,\,\,\,\,\,\,\,\,\,\,\,\,\,\,\,\,\,\,\,(4.14) \nnb\\
x_1^2+x_2^2+y_1^2+y_2^2+z_2^2=c_2,\,\,\,\,\,\,\,\,\,\,\,\,\,\,\,\,\,\,\,\,\,\,\,\,\,\,\,\,\,\,\,\,
\,\,\,\,\,\,\,\,\,\,\,\,\,\,\,\,\,\,\,\,\,\,\,(4.9) \nnb\end{array}\right.
\eea
Using the data of the neutrino mass squared differences, $\Delta m_{21}^2$,$\Delta m_{31}^2$, and mixing angles\cite{Esteban}, we have the values of $a_2$ and $w$: $a_2=5.579\times 10^{-5} eV^2, w=4.676$. The Eqs.(4.4-4.9) are secondary simultaneous equations in six variables. We found two particular solutions of the aboved equations. Their detailed expressions are given in Appendix. To see the order of elements, which are called as the matrix parameters for simplicity hereafter, in the matrix $M_\nu$ (Eq.(4.1)), we list numerical value results of the two solutions below.

For the case of $y_1=y_2$,  the matrix parameters are\footnote{The unity of the matrix parameters is eV. All quantities in the following have the unity $eV$ or $eV^2$.}\bea
a&=&-3.844\times10^{-3},\,\,\,b=(0.1716-3.751\mathrm{i})\times10^{-2},\,\,\,\nnb\\
d&=&-8.023\times10^{-3}(1+\mathrm{i}),\,\,\,e=1.150\times10^{-2}.\nnb
\eea

So it is straightforward to get one solution of matrix $M_\nu$  from Eq.(4.1):

\bea \hskip-1cm M_\nu&=&\footnotesize{\left(\begin{array}{lll}
-3.844\times10^{-3}&(0.1716-3.751\mathrm{i})\times10^{-2}&(-0.1716-3.751\mathrm{i})\times10^{-2}\\
(0.1716-3.751\mathrm{i})\times10^{-2}&-8.023\times10^{-3}(1+\mathrm{i})&
1.150\times10^{-2}\\
(-0.1716-3.751\mathrm{i})
\times10^{-2}&1.150\times10^{-2}&-8.023\times10^{-3}(1-\mathrm{i})
\end{array}\right)}.
\,\,\,\,\,\,\,\,\,\,\,\,\,\,\,\,\,\,\,\,\,\,\,\,\,\,\,(4.15)\nnb\eea

%\bea \hskip-1cm M_\nu&=&\footnotesize{\left(\begin{array}{lll}
%-3.844\times10^{-3}&(0.1716-3.751\mathrm{i})\times10^{-2}\\
%(0.1716-3.751\mathrm{i})\times10^{-2}&-8.023\times10^{-3}(1+\mathrm{i})\\
%(-0.1716-3.751\mathrm{i})\times10^{-2}&1.150\times10^{-2}\end{array}\right.}\nnb\\
%&&\footnotesize{\left.\begin{array}{c}
%(-0.1716-3.751\mathrm{i})\times10^{-2}\\
%1.150\times10^{-2}\\
%-8.023\times10^{-3}(1-\mathrm{i})\end{array}\right)}.
%\,\,\,\,\,\,\,\,\,\,\,\,\,\,\,\,\,\,\,\,\,\,\,\,\,\,\,\,\,\,\,\,\,\,\,\,\,\,\,\,\,\,\,\,\,\,\,(4.15)\nnb\eea

The another solution of mass matrix $M_\nu$ with $x_1=-x_2$ is:

\bea
\hskip-1cm
M_\nu&=&\footnotesize{\left(\begin{array}{lll}
-5.324\times10^{-2}&2.224\times10^{-4}(1-\mathrm{i})&-2.224\times10^{-4}(1-\mathrm{i})\\
2.224\times10^{-4}(1-\mathrm{i})&(0.1173-5.376\mathrm{i})\times10^{-2}&5.198\times10^{-4}\\
-2.224\times10^{-4}(1-\mathrm{i})&5.198\times10^{-4}&1.173\times10^{-3}+5.376\times10^{-2}\mathrm{i}
\end{array}\right)}.
\,\,\,\,\,\,\,\,\,\,\,\,\,\,\,\,\,\,\,\,\,\,\,\,\,\,\,\,\,\,\,\,\,(4.16)\nnb\eea

%\bea
%\hskip-1cm
%M_\nu&=&\footnotesize{\left(\begin{array}{lll}
%-5.324\times10^{-2}&2.224\times10^{-4}(1-\mathrm{i})\\
%2.224\times10^{-4}(1-\mathrm{i})&(0.1173-5.376\mathrm{i})\times10^{-2}\\
%-2.224\times10^{-4}(1-\mathrm{i})&5.198\times10^{-4}\end{array}\right.}\nnb\\
%&&\footnotesize{\left.\begin{array}{c}
%-2.224\times10^{-4}(1-\mathrm{i})\\
%5.198\times10^{-4}\\
%1.173\times10^{-3}+5.376\times10^{-2}\mathrm{i}\end{array}\right)}.
%\,\,\,\,\,\,\,\,\,\,\,\,\,\,\,\,\,\,\,\,\,\,\,\,\,\,\,\,\,\,\,\,\,\,\,\,\,\,\,\,(4.16)\nnb\eea

Because there are indeed five variables in two particular solutions, no free parameter ($c_1$ or $c_2$) remains in the solutions.

2)Model II\\

 In this model, one neutrino mass matrix element vanishes\cite{Yue}. So the mass matrix $M_\nu$ could be expressed as

\bea  M_\nu&=& \left(\begin{array}{ccc}
0&b&b^*\\
b&d&e\\
b^*&e&d^*\end{array}\right),
 \,\, \,\,\,\,\,\,\,\,\,\,\,\,\,\,\,\,\,\,\,\,\,\,\,\,\,\,\,\,\,\,\,\,\,\,\,\,\,\,\,\,\,\,\,\,\,\,\,\,\,\,\,\,\,\,\,\,\,\,\,\,\, (4.17) \nnb \eea
where $e$ is real.

Comparing with Eq.(3.1), we have $a=0,c=b^*,f=d^*$. Then the following relations are derived from Eq.(3.3)
\bea A&=&2|b|^2, \,\,\,\,B= bd^*+b^*e,\,\,\,\ C=be+b^*d=B^*,\,\,\,\,\nnb\\
D&=&|b|^2+|d|^2+e^2, \,\,\,\,E=b^2+2de,\,\,\,\, F=D.  \nnb\eea
Beginning with $C=B^*$, one could derive that ${\rm \Re}(B)={\rm \Re}(C), {\rm \Im}(B) =-{\rm \Im}(C)$
and considering Eq.(3.22), we have $\tan\theta_{23}=-1, \cos(2\theta_{23})=0.$
% Therefore, Eq.(3.9) changes into $\chi= 2\cos\theta_{23}{\rm \Re}(B)/(\cos\theta_{13}).$
Next, the relation $F=D$, $\cos(2\theta_{23})=0,$ and Eq.(3.16) lead to the result of $\cos \delta=0$, as it is expected. Then,
we have ${\rm \Re}(E) =\sin(\theta_{23}{\rm \Im}(C))/(\tan\theta_{13}\sin\delta).$ Consequently, we could get the following equations:\\
\bea
  &&x_1(y_1+z_2)+x_2y_2=b_1, \,\,\,\,\,\,\,\,\,\,\,\,\,\,\,\,\,\,\,\,\,\,\,\,\,\,\,\,\,\,\,\,\,\,\,\,\,\,\,\,\,\,\,\,\, \,  \,\,\,\,\, \,\,\, \,(4.18)\nnb\\
    &&x_2(z_2-y_1)+x_1y_2=b_2,  \,\,\,\,\,\,\, \,\,\,\,\, \,\,\, \,\,\,\,\,\,\,\,\,\,\,\,\,\,\,\,\,\, \,\,\,\,\,\,\,\,\,\,\,\, \,\,\,\,\,\,\,\,\,\,(4.19)\nnb\\
        &&x_1^2-x_2^2+2z_2y_1=a_1, \,\,\,\,\,\,\,\,\,\,\,\,\,\,\,\,\,\,\,\,\,\,\,\,\,\,\,\,\,\,\,\,\,\,\,\,\,\,\,\,\,\,\,\,\,\,\,\,\,\,\,\,\,\,\,\,\,\,\,\,\,\, (4.20)\nnb\\
        &&-2(x_1x_2+z_2y_2)=a_2,\,\,\, \,\,\, \,\,\, \,\,\,\,\,\,\,\,\,\,\,\,\,\,\,\,\,\, \,\,\,\,\,\,\,\,\,\,\,\,\,\, \,\,\,\,\,\,\,\,\,\,\,\,\,\,\,\,\,\,\,\,(4.21)\nnb\\
    &&y^2_1+y^2_2=c_2-c_1/2,\,\,\,\,\,\,  \,\,\,  \,\,\,  \,\,\,  \,\,\,  \,\,\,\,  \,\,\,  \,\,\,\,\,\,\,\,\,\,\,\,\,\,\,\,\,\, \,\,\,\,\,\,\,\,\,\,\,\,\,\,\,\,\,\,\,\,\,\,\,(4.22)\nnb
\eea
where $x_1$ and $x_2$ satisfy $x_1^2+x_2^2=c_1/2.$

Two particular solutions of the mass matrix $M_\nu$ are derived and listed in Appendix. %Here comes the numerical results of the solutions:
We give numerical results of the solutions in the following.

When $b_1=b_2$ is set up, one has the expression of $M_\nu$:
\begin{eqnarray}
\hskip-1cm
M_\nu&=&\footnotesize{\left(\begin{array}{lll}0&
(3.486-0.08753\mathrm{i})\times 10^{-2}&(3.486+0.08753\mathrm{i})\times 10^{-2} \\
(3.486-0.08753\mathrm{i})\times 10^{-2} &7.309\times 10^{-3}(1+\mathrm{i})&3.580\times 10^{-4}\\
(3.486+0.08753\mathrm{i})\times 10^{-2}&3.580\times 10^{-4}&7.309\times 10^{-3}(1-\mathrm{i})\end{array}\right)}.(4.23)\nnb\end{eqnarray}

%\bea
%\hskip-1cm
%M_\nu&=&\footnotesize{\left(\begin{array}{lll}
%(3.486-0.08753\mathrm{i})\times 10^{-2} \\
%(3.486-0.08753\mathrm{i})\times 10^{-2} &7.309\times 10^{-3}(1+\mathrm{i})\\
%(3.486+0.08753\mathrm{i})\times 10^{-2}&3.580\times 10^{-4}\end{array}\right.}\nnb\\
%&&\footnotesize{\left.\begin{array}{c}
%(3.486+0.08753\mathrm{i})\times 10^{-2}\\
%3.580\times 10^{-4}\\
%7.309\times 10^{-3}(1-\mathrm{i})\end{array}\right)}.
%\,\,\,\,\,\,\,\,\,\,\,\,\,\,\,\,\,\,\,\,\,\,\,\,\,\,\,\,\,\,\,\,\,\,\,\,\,\,\,\,\,\,\,\,\,\,\,\,\,\,(4.23)\nnb\eea

When $b_1=-b_2$ holds, one has
\begin{eqnarray}
\hskip-1cm
M_\nu&=&\footnotesize{\left(\begin{array}{lll}0&
(-3.493+0.07986\mathrm{i})\times10^{-2} &(-3.493-0.07986\mathrm{i})\times10^{-2}\\
(-3.493+0.07986\mathrm{i})\times10^{-2} &-7.301\times10^{-3}\mathrm{i}&7.301\times10^{-3}\\
(-3.493-0.07986\mathrm{i})\times10^{-2}&7.301\times10^{-3}&7.301\times10^{-3}\mathrm{i}
\end{array}\right)}.(4.24)\nnb\end{eqnarray}

Comparing with the result in Ref.\cite{Yue}, the order of numerical values of matrix elements is the same. Therefore, their model could be included into the kind of model that we investigated in this paper so that
 $\cos\delta=0$ must valid.

\section{Summary and Conclusions}

Using a very interesting equation inspired by an interesting condition \cite{Buras}, we have examined a kind of $3\times 3$ Hermitian matrix of which Real component and Imaginary part are commutative and shown an important conclusion that the kind of $3\times 3$ Hermitian matrix has only five real parameters. We have carried out diagonalization of the Hermitian Majorana neutrino mass matrix $M=M_\nu M_\nu^\dagger$ of which Real component and Imaginary part are commutative. Because the matrix elements of M have to satisfy four constraints we obtain in a model-independent way an important result, Dirac CP violation phase $\delta = \pm(\pi/2+n\pi), n=0,1,2...$, a prediction, atmospherical mixing angle would be in the following ranges: $\pi/4 < \theta_{23} < 3\pi/4$, or $-3\pi/4<\theta_{23} < - \pi/4$,
 and some sum rule for mixing angles which is in agreement with data of the current global fits\cite{Esteban,NuFIT} at 3 sigma level for both mass hierarchies. We have shown obviously that the mass matrix $M$ has only five real parameters and furthermore, only one free real parameter (A or D) if using the measured values of three mixing angles and mass squared differences as inputs. We have given matrix elements of neutrino mass matrix $M_\nu$ explicitly in some specific examples which should be helpful for investigating theoretical models in neutrino physics.

 In the Ref.\cite{Tang} authors examine the physics reach of the proposed medium baseline muon decay experiment MOMENT. To reach the precision of $\delta_{CP}$ at $10^\circ$ or better at $1\sigma$ confidence level, they find it sufficient to combine the data of MOMENT, DUNE and T2HK. The authors in the Ref.\cite{Tek} examine confidence intervals on $\delta_{CP}$ from the T2K plus Reactors fit in both the normal (NO) and inverted (IO) ordering. Their results (see Fig.4 in the reference) show that
the values of $\delta_{CP}$ are in (-2.42, -1.2) interval at  the $68.27 \% $ confidence in the normal ordering and there are no values in the inverted ordering inside the $68.27  \%$ confidence interval.

% the physics reach of the proposed medium baseline muon decay experiment MOMENT. To reach the precision of $\delta_{CP}$ at $10^\circ$ or better at $1\sigma$ confidence level, they %find it sufficient to combine the data of MOMENT, DUNE and T2HK. Besides, the study by Ushak Rahaman and Soebur Razzaque has suggested that the $NO\nu A$ near detector data favor %the case of non-unitary mixing
%of three neutrinos and non-unitary mixing favors $\delta_{CP} \approx +\frac{\pi}{2}( -\frac{\pi}{2})$ for normal(inverted) mass hierarchy for a combined analysis of near and far %detector data\cite{ur}.
It is expected that the precision measurement of $\delta_{CP}$ in the near future will help people to research the kind of neutrino models that we examined in the paper. \\

\section*{Acknowledgement}
This research was supported in part by the Natural Science Foundation of China under grant numbers (No. 11847612 and No. 11875306).\\
\\

\center{\Large{Appendix}}
\\
%\vskip0.5cm\setlength{\mathindent}{0.2cm}

   This section gives the details of the Majorana neutrino mass matrix forms in Model I and Model II with parameters $a_2=5.579\times10^{-5}eV^2, w=4.676$.
\vskip0.5cm

1.The Model I

In this model, we could get the Eqs.(4.9-4.14) from Eqs.(4.4-4.9) with $x_1={\rm \Re}(b), x_2={\rm \Im}(b), y_1={\rm \Re}(d), y_2={\rm \Im}(d)$, $z_1=a, z_2=e$ and $a_1={\rm \Re}(E), a_2={\rm \Im}(E), b_1={\rm \Re}(B), b_2={\rm \Im}(B)$, $c_1=A, c_2=D$, and relations of $a_1=\pm w^2a_2,
w=1/(t_{13} \sqrt{2}), b_2=w a_2.$\\

Thus, two solutions could be obtained from Eqs.(4.9-4.14), one case is to satisfy the condition of $y_1=y_2$, and the other is that the condition of $x_1=-x_2$ holds.
\vskip0.5cm
Case (1):$y_1=y_2$

\bea
x_1&=&\sqrt{\frac{a_2(1+w^2)}{w^4-2w^2-1}}
\,\,\,\,\,\,\,\,\,\,\,\,\,\,\,\,\,\,\,\,\,\,\,
\,\,\,\,\,\,\,\,\,\,\,\,\,\,\,\,\,\,\,\,\,\,\,\,\,\,\,\,
\,\,\,\,\,\,\,\,\,\,\,\,\,\,\,\,\,\,\,\,\,\,\,\,\,(A.1)\nnb\\
x_2&=&-w^2\sqrt{\frac{a_2(1+w^2)}{w^4-2w^2-1}},\,\,\,\,\,\,\,\,\,\,\,\,\,\,\,\,\,\,\,\,\,\,\,\,\,\,\,\,\,\,\,\,\,\,\,\,\,\,\,\,\,\,\,\,\,\,\,\,\,\,\,\,\,\,\,\,\,\,\,\,\,\,\,\,(A.2)\nnb\\
y_1&=&-w\sqrt{\frac{a_2(1+w^2)}{w^4-2w^2-1}},\,\,\,\,\,
\,\,\,\,\,\,\,\,\,\,\,\,\,\,\,\,\,\,\,\,\,\,\,\,\,\,\,\,\,\,\,\,\,\,\,\,\,\,\,\,\,\,\,\,\,\,\,\,\,\,\,\,\,\,\,\,\,\,\,\,\,\,(A.3)\nnb\\
y_2&=&y_1,\,\,\,\,\,\,\,\,\,\,\,\,\,\,\,\,\,\,\,\,\,\,\,\,\,\,\,\,\,\,\,\,
\,\,\,\,\,\,\,\,\,\,\,\,\,\,\,\,\,\,\,\,\,\,\,\,\,\,\,\,\,\,\,
\,\,\,\,\,\,\,\,\,\,\,\,\,\,\,\,\,\,\,\,\,\,\,\,\,\,\,\,\,\,\,\,\,\,
\,\,\,\,\,\,\,\,\,\,\,\,(A.4)\nnb\\
z_1&=&\sqrt{\frac{a_2(1+w^2)}{w^4-2w^2-1}}\cdot \nnb\\
&&\biggl[\frac{w(1+w^4)}{(1+w^2)} +\frac{(1+w^4)(1+4w^2+3w^4-2w^6)}{2w(w^4-2w^2-1)(1+w^2)}\biggl],
\,\,\,\,(A.5)\nnb\\
z_2&=&-\frac{\sqrt{\frac{a_2(1+w^2)}{w^4-2w^2-1}}}{2w(1+w^2)}
\biggl(1-3w^4\biggl),\,\,\,\,\,\,\,\,\,\,\,\,\,\,\,\,\,\,\,\,\,\,\,\,\,\,\,\,\,\,\,\,\,\,\,\,\,\,\,\,\,\,\,\,\,\,\,\,\,\,\,\,\,\,\,\,\,\,\,\,(A.6)\nnb\eea

   With $a_2=5.579\times10^{-5}, w=4.676$, the numerical results could be listed:
\bea
x_1&=&1.716\times10^{-3},\,\,\,x_2=-3.751\times10^{-2},\,\,\,\nnb\\
y_1&=&y_2=-8.023\times10^{-3},\,\,\,\nnb\\
z_1&=&-3.844\times10^{-3},\,\,\,z_2=1.15\times10^{-2}.\nnb\,\,\,\,\,\,\,\,\,\,\,\,\,\,\,\,\,\,\,\,\,\,\,\,\,\,\,\,\,\,\,\,\,\,(A.7)\eea

According to the relations of
$x_1={\rm \Re}(b), x_2={\rm \Im}(b)$, $y_1={\rm \Re}(d)$, $y_2={\rm \Im}(d)$, $z_1=a, z_2=e$,
one could calculate the matrix parameters of
$a, b, d, e$:
\bea
a&=&-3.844\times10^{-3},\,\,\,b=1.716\times10^{-3}-3.751\times10^{-2}\mathrm{i},\,\,\,\nnb\\
d&=&-8.023\times10^{-3}(1+\mathrm{i}),\,\,\,e=1.15\times10^{-2}.\nnb\,\,\,\,\,\,\,\,\,\,\,\,\,\,\,\,\,\,\,\,\,\,\,\,(A.8)
\eea
 The form of mass matrix $M_\nu$ could be written as (4.15) in section 4.
\vskip0.5cm
Case (2):$x_1=-x_2$
\vskip0.5cm
Similarly, we could get the following expressions:
\bea
x_1&=&-x_2,\,\,\,\,\,\,\,\,\, \,\,\,\,\,\,\,\,\,\,\,\,\,\,\,\,\,\,\,\,\,\,\,\,\,\,\, \,\,\,\,\,\,\,\,\, \,\,\,\,\,\,\,\,\,\,\,\,\,\,\,\,\,\,\,\,\,\,\,\,\,\,
\,\,\,\,\,\,\,\,\,\,\,\,\,\,\,\,\,\,\,\,\,\,\,\,\,\,\,\,\,\,\,\,\,\,\,\,(A.9)\nnb\\
x_2&=&-\frac{b_2}{a_1^2+4b_2^2}\cdot M,\,\,\,\,\,\,\,\,\,\,\,\,\,\,\,\,\,\,\,\,\,\,\,\,\,\,\,\,\,\,\,\,\,\,\,\,\,\,\,\,\,\,\,\,\,\,\,\,\,\,\,\,\,\,\,\,\,\,\,\,\,\,\,\,\,\,\,\,\,\,\,\,\,\,\,\,\,\,\,\,(A.10)\nnb \\
y_1&=&-\frac{b_2}{x_2},\,\,\,\,\,\,\,\,\,\,\,\,\,\,\,\,\,\,\,\,\,\,\,\,\,\,\,\,\,\,\,\,\,\,\,\,\,\,\,\,\,
\,\,\,\,\,\,\,\,\,\,\,\,\,\,\,\,\,\,\,\,\,\,\,\,\,\,\,\,\,\,\,\,\,\,\,\,\,\,\,\,\,\,\,\,\,\,\,\,\,\,\,\,\,\,\,\,\,\,\,\,\,\,\,\,\,\,(A.11)\nnb\\
y_2&=&
\biggl((a_1^2+4b_2^2)\cdot M\biggl)^{-1}\cdot\nnb\\
&&\biggl[4b_2^2\sqrt{(2a_2+c_1)(a_1^2c_1-8a_2b_2^2)}-a_1^3a_2-4a_1b_2^2\biggl],\,\,(A.12)\nnb\\
z_1&=&
\biggl[a_1^3c_1+4b_2^2\sqrt{(2a_2+c_1)(a_1^2c_1-8a_2b_2^2)}\nnb\\&&
-a_1\biggl(16a_2b_2^2+4b_2^2c_1+a_1\sqrt{(2a_2+c_1)(a_1^2c_1-8a_2b_2^2)}\biggl)
\biggl]\cdot\nnb\\&&
\biggl((a_1^2+4b_2^2)\cdot M\biggl)^{-1},\,\,\,\,\,\,\,\,\,\,\,\,\,\,\,\,\,\,\,\,\,\,\,\,\,\,\,\,\,\,\,\,\,\,\,\,\,\,\,\,\,\,\,\,\,\,\,\,\,\,\,\,\,\,\,\,\,\,\,\,\,\,\,\,\,\,\,\,\,\,\,\,(A.13)\nnb\\
z_2&=&\frac{a_1\cdot M}{2(a_1^2+4b_2^2))},\,\,\,\,\,\,\,\,\,\,\,\,\,\,\,\,\,\,\,\,\,\,\,\,\,\,\,\,\,\,\,\,\,\,\,\,\,\,\,\,\,\,\,\,\,\,\,\,\,\,\,\,\,\,\,\,\,\,\,\,
\,\,\,\,\,\,\,\,\,\,\,\,\,\,\,\,\,\,\,\,\,\,\,\,\,\,(A.14)\nnb\\
%\eea
%$where$\\
%\bea
M&=&\biggl[2a_1^2(a_2+c_1)\nnb\\
&&-2\biggl(4a_2b_2^2+a_1\sqrt{(2a_2+c_1)(a_1^2c_1-8a_2b_2^2)}\biggl)\biggl]^{-\frac{1}{2}},\nnb\\
%\eea\bea
c_1&=&(1 + 12 w^2 + 42 w^4 + 72 w^6 + 81 w^8 + 28 w^{10} + 8 w^{12})\nnb\\
&&(a_2+2b_2w+b_2w^3)/[4w^2(1+w^2)^3(w^4-2w^2-1)]\nnb\\
&-&\frac{a_2 + 6 b_2 w + 14 b_2w^3 + 16 b_2 w^5 + 5 b_2w^7 + 2 b_2w^9}{(1 + w^2) (-1 - 2 w^2 + w^4)}.\nnb
\eea
 %[²åÈëc1µÄ±í´ïÊ½]
With inputs of $w$ and $a_2$, we get $c_1=2.835\times10^{-3}$
and the numerical results are given %with $c_1=2.835\times10^{-3}$:
\bea
x_1&=&2.224\times10^{-4},\,\,\,
x_2=-2.224\times10^{-4},\,\,\,\nnb\\
y_1&=&1.173\times10^{-3},\,\,\,y_2=-5.376\times10^{-2},\,\,\,\nnb\\
z_1&=&-5.324\times10^{-2},\,\,\,z_2=5.198\times10^{-4}.\nnb
\,\,\,\,\,\,\,\,\,\,\,\,\,\,\,\,\,\,\,\,\,\,\,\,\,(A.15)
\eea

 The matrix parameters of $a, b, d, e$ are
\bea
a&=&-5.324\times10^{-2},\,\,\,b=2.224\times10^{-4}(1-\mathrm{i}),\nnb\\
d&=&10^{-2}\times(0.1173-5.376\mathrm{i}),\,\,\,e=5.198\times10^{-4}.\nnb\,\,\,(A.16)
\eea

The mass matrix $M_\nu$ could see Eq.(4.16) in section 4.
\vskip0.5cm
2.The Model II
\vskip0.5cm
In this model, one neutrino mass matrix element vanishes.
According to the analysis in section 4, one could solve Eqs.(4.18-4.22) and get two solutions with
\bea
a_1&=&{\rm \Re}(E), a_2={\rm \Im}(E), b_1={\rm \Re}(B), b_2={\rm \Im}(B),\nnb\\
c_1&=&A,c_2=D, x_1={\rm \Re}(b), x_2={\rm \Im}(b),\nnb\\ y_1&=&{\rm \Re}(d), y_2={\rm \Im}(d),z_2=e,z_1=a=0,\nnb\\ w&=&1/(t_{13}\sqrt{2}),a_1=w b_2, b_2=w a_2.\nnb\eea

Here two cases are discussed with $b_1=\pm b_2$ established.

\vskip0.5cm Case (1):$b_1=b_2$

\bea
x_1&=&u-\sqrt{\frac{u^2-a_1+a_2}{2}},
 \,\,\,\,\,\,\,\,\,\,\,\,\,\,\,\,\,\,\,\,\,\,\,\,\,\,\,\,\,\,\,\,\,\,\,\,\,\,\,\,\,\,\,\,\,\,\,\,\,\,\,\,\,\,\,\,\,(A.17)\nnb\\
x_2&=&-\sqrt{\frac{u^2-a_1+a_2}{2}},\,\,\,\,\,\,\,\, \,\,\,\,\,\,\,\,\,\,\,\,\,\,\,\,\,\,\,\,\,\,\,\,\,\,\,\,\,\,\,\,\,\,\,\,\,\,\,\,\,\,\,\,\,\,\,\,\,\,\,\,\,\,\,(A.18)\nnb\\
y_1&=&\frac{b_2u}{a_1+a_2}, \,\,\,\,\,\,\,\,\,\,\, \,\,\,\,\,\,\,\,\,\,\, \,\,\,\,\,\,\,\,\,\,\,\,\,\,\,\,\,\,\,\,\,\,\,\,\,\,\,\,\,\,\,\,\,\,\,\,\,\,\,\,\,\,\,\,\,\,\,\,\,\,\,\,\,\,\,\,\,\,\,\,\,\,\,(A.19)\nnb\\
 y_2&=&y_1,\,\,\,\,\,\,\,\,\, \,\,\,\,\,\,\,\,\,\,\,\,\,\,\,\,\, \,\,\,\,\,\,\,\,\,\,\,\,\,\,\,\,\,\,\,\,\,\,\,\,\,\,\,\,\,\,\,\,\,\,\,\,\,\,\,\,\,\,\,\,\,\,\,\,\,\,
\,\,\,\,\,\,\,\,\,\,\,\,\,\,\, \,\,\,\,\,\,\,\,(A.20)\nnb\\
 z_2&=&\frac{2b_2\sqrt{\frac{u^2-a_1+a_2}{2}}}{a_1+a_2}, \,\,\,\,\,\,\,\,\,\,\,\,\,\,\,\, \,\,\,\,\,\,\,\,\,\,\, \,\,\,\,\,\,\,\,\,\,\,\,\, \,\,\,\,\,\,\,\,\,\,\,\,\,\,\,\,\,\,\,\,\,\,\,\,\,\,\,\,\,\,(A.21)
\nnb\\
u&=&\sqrt{\frac{M_1+ \sqrt{M_2}}{M_3}},\nnb
\\
M_1&=&a_1^4 a_2 + a_2^5 - 4 a_2^3 b_2^2 + 4 a_2 b_2^4 + 4 a_1^3 (a_2^2 - b_2^2)
\nnb\\ &&+
 6 a_1^2 (a_2^3 - 2 a_2 b_2^2) +
 4 a_1 (a_2^4 - 3 a_2^2 b_2^2 +
    b_2^4), \nnb\\
M_2&=&(a_1 + a_2)^2 (a_1^2 + 2 a_1 a_2 + a_2^2 -
      2 b_2^2)^2 \nnb\\&&\cdot(a_1^4 + 2 a_1^3 a_2 +
      2 a_1^2 (a_2^2 - 2 b_2^2) + (a_2^2 - 2 b_2^2)^2\nnb\\&& +
      2 a_1 (a_2^3 - 4 a_2 b_2^2)),\nnb\\
M_3&=&a_1^4 + 4 a_1^3 a_2 + a_2^4 - 8 a_2^2 b_2^2 + 8 b_2^4\nnb\\&& +
  a_1^2 (6 a_2^2 - 8 b_2^2) + 4 a_1 (a_2^3 - 4 a_2 b_2^2).\nnb\eea

Inputting the values of $w$ and $a_2$, we have $u=3.574\times 10^{-2}$ and then the numerical solution is:\bea
x_2&=&-8.753\times 10^{-4},\,\,\,x_1=3.486\times 10^{-2},\nnb \\
y_2&=&y_1=7.309\times 10^{-3},\,\,\,\,\,\nnb \\
z_2&=&3.580\times 10^{-4}.\nnb\,\,\,\,\,\,\,\,\,\,\,\,\,\,\,\,\,\,\,\,\,\,\,\,\,\,\,\,\,\,\,\,\,\,\,\,\,\,\,\,\,\,\,\,\,\,\,\,\,\,\,\,\,\,\,\,\,\,\,\,\,\,\,\,\,\,\,\,\,\,\,\,\,\,\,\,\,(A.22)
\eea
Consequently the results of mass matrix parameters are
\bea b&=&3.486\times 10^{-2}-8.753\times 10^{-4}\mathrm{i},\nnb\\
d&=&7.309\times 10^{-3}(1+\mathrm{i}),\nnb\\
e&=&3.580\times 10^{-4}.\nnb\,\,\,\,\,\,\,\,\,\,\,\,\,\,\,\,\,\,\,\,\,\,\,\,\,\,\,\,\,\,\,\,\,\,\,\,\,\,\,\,\,\,\,\,\,\,\,\,\,\,\,\,\,\,\,\,\,\,\,\,\,\,\,\,\,\,\,\,\,\,\,\,\,\,\,\,\,\,\,(A.23)
\eea

The mass matrix could be written as Eq.(4.23) in section 4.

\vskip0.5cm
Case (2): $b_1=-b_2$

In the the same way, the following solution could be gotten:
\bea
x_2&=&\sqrt{\frac{-a_1+\sqrt{a_1^2+a_2^2}}{2}},\,\,\,\,\,\,\,\,\,\,\,\,\,\,\,\,\,\,\,\,\,\,\,\,\,\,\,\,\,\,\,\,\,\,\,\,\,\,\,\,\,\,\,\,\,\,\,\,\,\,\,\,\,\,\,\,\,\,\,\,(A.24)\nnb\\
x_1&=&\sqrt{\frac{-a_1+\sqrt{a_1^2+a_2^2}}{2}}+u,\,\,\,\,\,\,\,\,\,\,\,\,\,\,\,\,\,\,\,\,\,\,\,\,\,\,\,\,\,\,\,\,\,\,\,\,\,\,\,\,\,\,\,\,\,\,\,\,\,\,\,(A.25)\nnb\\
y_1&=&0,\,\,\,\,\,\,\,\,\,\,\,\,\,\,\,\,\,\,\,\,\,\,\,\,\,\,\,\,\,\,\,\,\,\,\,\,\,\,\,\,\,\,\,\,\,\,\,\,\,\,\,\,\,\,\,\,\,\,\,\,\,\,\,\,\,\,\,\,\,\,\,\,\,\,\,\,\,\,\,\,\,\,\,\,\,\,\,\,\,\,\,\,\,\,\,\,\,\,\,\,\,\,\,\,\,(A.26)\nnb\\
y_2&=&\frac{b_2}{u},\,\,\,\,\,\,\,\,\,\,\,\,\,\,\,\,\,\,\,\,\,\,\,\,\,\,\,\,\,\,\,\,\,\,\,\,\,\,\,\,\,\,\,\,\,\,\,\,\,\,\,\,\,\,\,\,\,\,\,\,\,\,\,\,\,\,\,\,\,\,\,\,\,\,\,\,\,\,\,\,\,\,\,\,\,\,\,\,\,\,\,\,\,\,\,\,\,\,\,\,\,(A.27)\nnb\\
z_2&=&-y_2,\,\,\,\,\,\,\,\,\,\,\,\,\,\,\,\,\,\,\,\,\,\,\,\,\,\,\,\,\,\,\,\,\,\,\,\,\,\,\,\,\,\,\,\,\,\,\,\,\,\,\,\,\,\,\,\,\,\,\,\,\,\,\,\,\,\,\,\,\,\,\,\,\,\,\,\,\,\,\,\,\,\,\,\,\,\,\,\,\,\,\,\,\,\,\,\,\,\,\,(A.28)\nnb\\
u&=&\frac{a_1-a_2-\sqrt{a_1^2+a_2^2}}{\sqrt{2(\sqrt{a_1^2+a_2^2}-a_1)}}.\nnb \eea

With the inputs of $w=4.676, a_2=5.579\times10^{-5}$, one has $b_2=2.609\times10^{-4}, a_1=1.22\times10^{-3},u=-3.573\times10^{-2}$  and the numerical results of the solution are given:
\bea
x_1&=&-3.493\times10^{-2},\,\,\,x_2=7.986\times10^{-4},\nnb\\
y_1&=&0,\,\,\,
y_2=-7.301\times10^{-3},\nnb\\
z_2&=&7.301\times10^{-3}.\nnb\,\,\,\,\,\,\,\,\,\,\,\,\,\,\,\,\,\,\,\,\,\,\,\,\,\,\,\,\,\,\,\,\,\,\,\,\,\,\,\,\,\,\,\,\,\,\,\,\,\,\,\,\,\,\,\,\,\,\,\,\,\,\,\,\,\,\,\,\,\,\,\,\,\,\,\,\,(A.29)
\eea

Thus, the numerical values of mass matrix parameters are presented
\bea b&=&-3.493\times10^{-2}+7.986\times10^{-4}\mathrm{i},\nnb\\
d&=&-7.301\times10^{-3}\mathrm{i},\nnb\\
e&=&7.301\times10^{-3}.\nnb\,\,\,\,\,\,\,\,\,\,\,\,\,\,\,\,\,\,\,\,\,\,\,\,\,\,\,\,\,\,\,\,\,\,\,\,\,\,\,\,\,\,\,\,\,\,\,\,\,\,\,\,\,\,\,\,\,\,\,\,\,\,\,\,\,\,\,\,\,\,\,\,\,\,\,\,\,\,(A.30)
\eea

The mass matrix $M_\nu$ could be written as Eq.(4.24) in section 4.


\section*{References}
\begin{thebibliography}{999}
%\bibitem{PMNS}
%Particle Data Group, K. Nakamura {\it et al.}, J. Phys. {\bf {G 37}}, (2010) 075021;
%PMNS matrix was been introduced in B. Pontecorvo, JETP(USSR), {\bf {34}}, (1958) 247; Zh. Eksp. Teor. Fiz. {\bf {53}}, (1967) 1717;
%Z. Maki, M. Nakagawa, S. Sakata, Prog. Theor. Phys. {\bf {28}}, (1962) 870.
%\bibitem{Esteban}Esteban, {\it et al.}, IJMPA, {\bf {09}}, (2020) 178, arXiv:2007.14792[hep-ph].
\bibitem{tbm}P. F. Harrison, D. H. Perkins and W. G. Scott, Phys. Lett. {\bf {B 530}}, (2002) 167, arXiv:0202074[hep-ph]; P. F. Harrison and W. G. Scott, Phys. Lett. {\bf {B 535}}, (2002) 163-169, arXiv:0203209[hep-ph]; Z. z. Xing, Phys. Lett. {\bf {B 533}}, (2002) 85-93, arXiv:0204049 [hep-ph]; J. Ganguly, R. S. Hunid, arXiv:2107.07275[hep-ph].
\bibitem{He} X. G. He and A. Zee, Phys. Lett. {\bf {B 645}}, (2007) 427, arXiv:0607163[hep-ph]; X. G. He and A. Zee, Phys. Rev. {\bf {D 84}}, (2011) 0530, arXiv:1106.4359[hep-ph]; Carl H. Albright, Werner Rodejohann, Eur. Phys. J. {\bf {C 62}}, (2009) 599-608, arXiv:0812.0436[hep-ph]; Carl H. Albright, Alexander Dueck, Werner Rodejohann, Eur. Phys. J. {\bf {C 70}}, (2010) 1099-1110, arXiv:1004.2798[hep-ph].
\bibitem{Xi}Paul H. Frampton, Sheldon L. Glashow and Danny Marfatia, Phys. Lett. {\bf {B 536}}, 79 (2002), arXiv:0201008[hep-ph]; Zhi-zhong Xing, Phys. Lett. {\bf {B 530}}, (2002) 159, arXiv:0201151[hep-ph]; Bipin R. Desai, D. P. Roy and Alexander R. Vaucher, Mod. Phys. Lett. {\bf {A 18}}, (2003) 1355, arXiv:0209035[hep-ph]; A. Merle, M. Singh, G. Ahuja and M. Gupta, PTEP 2016, no.12, (2016) 123B08, arXiv:1603.08083[hep-ph], D. Borah, M. Ghosh, S. Gupta and S. K. Raut, Phys. Rev. {\bf {D 96}}, (2017) 055017, arXiv:1706.02017[hep-ph].
\bibitem{van}L. Lavoura, Phys. Lett. {\bf {B 609}}, (2005) 317, arXiv:0411232[hep-ph]; E. I. Lashin and N. Chamoun, Phys. Rev. {\bf {D 78}}, (2008) 073002, arXiv:0708.2423[hep-ph]; E. I. Lashin, N. Chamoun, Phys. Rev. {\bf {D 80}}, (2009) 093004, arXiv:0909.2669[hep-ph]; W. Wang, Phys. Lett. {\bf {B 733}}, (2014) 320, Erratum:[Phys. Lett. {\bf {B 738}}, (2014) 524], arXiv:1401.3949[hep-ph]; W. Wang, Phys. Rev. {\bf {D 90}}, no. 3, (2014) 033014, arXiv:1402.6808[hep-ph].
\bibitem{eq}S. Dev, R. R. Gautam and L. Singh, Phys. Rev. {\bf {D 87}}, (2013) 073011, arXiv:1303.3092[hep-ph].
\bibitem{hy} S. Kaneko, H. Sawanaka and M. Tanimoto, JHEP 0508, (2005) 073, arXiv:0504074[hep-ph]; S. Dev, S. Verma and S. Gupta, Phys. Lett. {\bf {B 687}}, (2010) 53-56, arXiv:0909.3182[hep-ph]; J. Y. Liu and S. Zhou, Phys. Rev. {\bf {D 87}}, no.9, (2013) 093010, arXiv:1304.2334[hep-ph]; W. Wang, Eur. Phys. J. {\bf {C 73}}, (2013) 2551, arXiv:1306.3556[hep-ph]; S. Dev, R. R. Gautam and L. Singh, Phys. Rev. {\bf {D 88}}, (2013) 033008, arXiv:1306.4281[hep-ph]; S. Dev and D. Raj, Nucl. Phys. {\bf {B 957}}, (2020) 115081, arXiv:2006.12019[hep-ph].
\bibitem{king}S. F. King and C. C. Nishi, Phys. Lett. {\bf {B 785}}, (2018) 391, arXiv:1807.00023[hep-ph]; C. C. Nishi, ${\it et \,al.}$, JHEP {\bf {B 09}}, (2018) 042, arXiv:1806.07412[hep-ph].
\bibitem{RR}R. R. Gautam, Sanjeev Kumar, Phys. Lett. {\bf {B 820}}, (2021) 136504, arXiv:2107.01526[hep-ph].
\bibitem{mkb}Mitesh Kumar Behera, ${\it et \,al.}$, arXiv:2108.04066[hep-ph].
\bibitem{JD}J. D. Garc$\acute{i}$a, J. C. G$\acute{o}$mez-Izquierdo, arXiv:2108.00317[hep-ph].
\bibitem{ai}A. Ismael, ${\it et \,al.}$, Nucl. Phys. {\bf {B 971}}, (2021) 115541, arXiv:2102.08326[hep-ph].
\bibitem{yu}Yuta Hyodo, and Teruyuki Kitabayashi, Progress of Theoretical and Experimental Physics, {\bf {B 123}}, (2021) 08, arXiv:2105.08210[hep-ph].
\bibitem{edv}Eleonora Di Valentino, ${\it et \,al.}$, Phys. Rev.{\bf {D 104}}, (2021) 083504, arXiv:2106.15267[hep-ph].
%\bibitem{Xic}%R. Adhikari, G. Rajasekaran, Phys. Rev. {\bf {D 61}} (2000) 031301, hep-ph/9812361;
%Zhi-zhong Xing, Ye-Ling Zhou, Phys. Rev {\bf {D 88}}, (2013) 033002, arXiv:1305.5718; Zhi-zhong Xing, Physics Reports {\bf {854}} (2020) 1, 1909.09610 [hep-ph].
\bibitem{Xi1}Ce-ran Hu, ZHi-Zhong Xing, Nucl. Phys. {\bf {B 971}}, (2021) 115521, arXiv:2108.00986[hep-ph];  H.
Georgi and S. L. Glashow, Phys. Rev. {\bf {D 61}}, (2000) 097301, hep-ph/9808293; G. C. Branco et
al, Phys. Rev. Lett. {\bf {82}} (1999) 683, hep-ph/9810328;  R. Adhikari and G. Rajasekaran,
Phys. Rev. {\bf {D 61}}, (2000) 031301, hep-ph/9812361.
\bibitem{Xic}%R. Adhikari, G. Rajasekaran, Phys. Rev. {\bf {D 61}} (2000) 031301, hep-ph/9812361;
Zhi-zhong Xing, Ye-Ling Zhou, Phys. Rev {\bf {D 88}}, (2013) 033002, arXiv:1305.5718; Zhi-zhong Xing, Physics Reports {\bf {854}} (2020) 1, 1909.09610 [hep-ph].
\bibitem{babu}K. S. Babu, ${\it et \,al.}$, arXiv:2108.11961[hep-ph].
\bibitem{ab}A. Branca, ${\it et \,al.}$, Symmetry {\bf {13}}, 9, (2021) 1625, arXiv:2108.12212[hep-ph].
\bibitem{PMNS}
Particle Data Group, K. Nakamura {\it et al.}, J. Phys. {\bf {G 37}}, (2010) 075021;
PMNS matrix was been introduced in B. Pontecorvo, Zh. Eksp. Teor. Fiz., {\bf {34}} (1958) 247, (Sov. Phys. JETP, {\bf {7}}, (1958) 172); Zh. Eksp. Teor. Fiz. {\bf {53}}, (1967) 1717, (Sov. Phys. JETP, {\bf {26}}, (1968) 984);
Z. Maki, M. Nakagawa, S. Sakata, Prog. Theor. Phys. {\bf {28}}, (1962) 870.
\bibitem{Esteban}Ivan Esteban, {\it et al.},  JHEP, {\bf {09}} (2020) 178, arXiv:2007.14792[hep-ph].
\bibitem{NuFIT}NuFIT, http://www.nu-fit.org/?q=node/8; NuFIT 5.1 (2021), www.nu-fit.org.
\bibitem{Singh}M. Singh, Prog Theor.Exp.Phys., {\bf {B04}} (2020) 093, arXiv:2006.09687[hep-ph].
\bibitem{Duarah}Chandan Duarah, arXiv:1905.07910[hep-ph].
\bibitem{Aizawa}Ichiro Aizawa, Masaki Yasu$\grave{e}$, Phys. Lett. {\bf {B 607}}, (2005) 267, arXiv:0409331[hep-ph];
\bibitem{xy}Xing Zhizhong, Ye-Ling Zhou, Phys. Lett. {\bf {B 693}}, (2010) 584, arXiv:1008.4906[hep-ph].
%\bibitem{He16}Junxing Pan, Jin Sun, and Xiao-Gang He,  Int. J. Mod. Phys.  {\bf {A 34}}, (2020) 35, 1950235, arXiv:1910.06688[hep-ph].
\bibitem{Jarlskog}C. Jarlskog, Phys. Rev. Lett. {\bf {55}}, (1985) 1039; Z.Z. Xing, Phys. Lett. {\bf {B 679}}, (2009) 111; and references therein.
\bibitem{gan}A. Gando et al. (KamLAND-Zen), Phys. Rev. Lett. {\bf {117}}, 082503 (2016), [Addendum:
Phys.Rev.Lett. 117, 109903 (2016)], arXiv:1605.02889 [hep-ex].\bibitem{Scott}
P. F. Harrison, W. G. Scott, Phys. Lett. {\bf {B 547}}, (2002) 219, arXiv:0210197[hep-ph].
\bibitem{Babu}
 K. S. Babu, E. Ma and J. W. F. Valle, Phys. Lett. {\bf {B 552}}, (2003) 207, arXiv:0206292[hep-ph].
\bibitem{Ma}
  E. Ma, Phys. Lett. {\bf {B 583}}, (2004) 157, arXiv:0308282[hep-ph].
\bibitem{Grimus}
   W. Grimus, L. Lavoura, Phys. Lett. {\bf {B 579}}, (2004) 113, arXiv:0305309[hep-ph].
\bibitem{he} X.-G. He, Chin. J. Phys. {\bf {53}}, (2015) 100101, arXiv:1504.01560v3[hep-ph].
\bibitem{Fukuyama}T. Fukuyama, PTEP {\bf {2017}}, (2017) 3, 033B11, arXiv:1701.04985v1[hep-ph].
\bibitem{xing}Z.-Z. Xing, Z.-H. Zhao, Rep. Prog. Phys. {\bf {79}}, (2016)  7:076201, arXiv:1512.04207[hep-ph].
\bibitem{Popov}E. Ma, A. Natale, O. Popov, Phys. Lett. {\bf {B 746}}, (2015) 114, arXiv:1502.08023[hep-ph].
%\bibitem{kn} S. F. King and C. C. Nishi, %Mu-tau symmetry and the Littlest Seesaw,
%Phys. Lett. {\bf {B 785}}, (2018) 391, arXiv:1807.00023.
%\bibitem{nss} C. C. Nishi, B. L. S¡äanchez-Vega and G. Souza Silva, %¦Ì¦Ó reflection symmetry
%with a high scale texture-zero,
JHEP {\bf {09}} (2018) 042, arXiv:1806.07412.
\bibitem{Yue}Zhi-Cheng Liu, Chong-Xing Yue and Zhen-hua Zhao, Phys. Rev. {\bf {D 99}}, (2019) 075034, arXiv:1808.06837[hep-ph].
\bibitem{Buras} G. C. Branco, A. J. Buras, and J. M. Gerard, Nucl. Phys. {\bf {B 259}}, (1985) 306.
\bibitem{Tang}Jian Tang, Sampsa Vihonen, JHEP, {\bf {2019}}, (2019) 12:130, arXiv:1909.01548[hep-ph].
%\bibitem{ur}Ushak Rahaman and Soebur Razzaque, arXiv:2108.11783[hep-ph].
%\bibitem{Buras} G. C. Branco, A. J. Buras, and J. M. Gerard, Nucl. Phys. {\bf {B 259}}, (1985) 306.
\bibitem{Tek}The T2K Collaboration, Nature {\bf {580}}, (2020) 339, arXiv:1910.03887[hep-ex].
\end{thebibliography}
\end{document}